\documentclass[
  aip,%
  amsmath,%
  amssymb,%
  reprint,%
]{revtex4-1}

\usepackage[T1]{fontenc}
\usepackage[utf8]{inputenc}
\usepackage{algorithm}
\usepackage{algpseudocode}
\usepackage{amsmath}
\usepackage{amssymb}
\usepackage[english]{babel}
\usepackage{booktabs}
\usepackage[acronym]{glossaries}
\usepackage[%
  debug,
  hidelinks,
  pdfauthor={Fabian Köhler},
  pdfusetitle,
  pdfversion=1.9,
  unicode=true,
]{hyperref}
\usepackage{graphicx}
\usepackage{todonotes}
\usepackage{physics}
\usepackage{placeins}
\usepackage{siunitx}
\usepackage[caption=false]{subfig}
\usepackage{xfrac}
\usepackage{xparse}
\usepackage{bm}
\usepackage{xcolor}

\let\vec\bm

\ExplSyntaxOn{}
\DeclareDocumentCommand\diff{o}{\IfNoValueTF{#1}{\mathrm{d}}{\mathrm{d}^{#1}}}
\ExplSyntaxOff{}

\begin{document}
\newacronym[longplural={equations of motion},shortplural={EoMs}]{eom}{EoM}{equation of motion}
\newacronym{bec}{BEC}{Bose-Einstein condensate}
\newacronym{dmrg}{DMRG}{density matrix renormalization group}
\newacronym{ec}{EC}{energy criterion}
\newacronym{gpe}{GPE}{Gross-Pitaevskii equation}
\newacronym{mctdhb}{MCTDHB}{multi-configuration time-dependent Hartree method for bosons}
\newacronym{mctdhf}{MCTDHF}{multi-configuration time-dependent Hartree-Fock}
\newacronym{mctdh}{MCTDH}{multi-configuration time-dependent Hartree}
\newacronym{ml-mctdhx}{ML-MCTDHX}{multi-configuration time-dependent Hartree method for mixtures}
\newacronym{mc}{MC}{magnitude criterion}
\newacronym{qmc}{QMC}{Quantum Monte Carlo}
\newacronym{ras}{RAS}{restriced-active-space}
\newacronym{spf}{SPF}{single-particle function}

\title{Dynamical Pruning of the Non-Equilibrium Quantum Dynamics of Trapped Ultracold Bosons}

\date{April 21, 2019}

\author{F.~Köhler}
\affiliation{Center for Optical Quantum Technologies, Department of Physics, University of Hamburg, Luruper Chaussee 149, 22761 Hamburg, Germany}
\affiliation{Hamburg Center for Ultrafast Imaging, University of Hamburg, Luruper Chaussee 149, 22761 Hamburg, Germany}
\author{K.~Keiler}
\affiliation{Center for Optical Quantum Technologies, Department of Physics, University of Hamburg, Luruper Chaussee 149, 22761 Hamburg, Germany}
\author{S.~I.~Mistakidis}
\affiliation{Center for Optical Quantum Technologies, Department of Physics, University of Hamburg, Luruper Chaussee 149, 22761 Hamburg, Germany}
\author{H.-D.~Meyer}
\affiliation{Theoretische Chemie, Physikalisch-Chemisches Institut, Universität Heidelberg, Im Neuenheimer Feld 229, 69120 Heidelberg, Germany}
\author{P.~Schmelcher}
\affiliation{Center for Optical Quantum Technologies, Department of Physics, University of Hamburg, Luruper Chaussee 149, 22761 Hamburg, Germany}
\affiliation{Hamburg Center for Ultrafast Imaging, University of Hamburg, Luruper Chaussee 149, 22761 Hamburg, Germany}

\begin{abstract}
  The investigation of the nonequilibrium quantum dynamics of bosonic many-body systems is very challenging due to the excessively growing Hilbert space and poses a major problem for their theoretical description and simulation.
  We present a novel dynamical pruning approach in the framework of the \gls{mctdhb} to tackle this issue by dynamically detecting the most relevant number states of the underlying physical system and modifying the many-body Hamiltonian accordingly.
  We discuss two different number state selection criteria as well as two different ways to modify the Hamiltonian.
  Our scheme regularly re-evaluates the number state selection in order to dynamically adapt to the time evolution of the system.
  To benchmark our methodology, we study the nonequilibrium dynamics of bosonic particles confined in either an optical lattice or in a double-well potential.
  It is shown that our approach reproduces the unpruned \gls{mctdhb} results accurately while yielding a significant reduction of the simulation time.
  The speedup is particularly pronounced in the case of the optical lattice.
\end{abstract}
\maketitle{}

\section{\label{sec:introduction}Introduction}
Ever since the first realizations of \glspl{bec}~\cite{anderson1995,bradley1995,davis1995}, ultracold atomic gases attracted a lot of interest both from the experimental and the theoretical side.
Their tunability and almost perfect isolation from the environment render such systems ideal candidates to simulate a variety of quantum many-body systems~\cite{bloch2008,bloch2012,polkovnikov2011}.
Due to experimental advancements ensembles of ultracold atoms with a controlled number of particles~\cite{serwane2011,kaufman2014} can be realized in arbitrarily shaped confining potentials~\cite{henderson2009} such as optical lattices~\cite{jaksch1998,bloch2005}, harmonic~\cite{chu1986} or ring traps~\cite{morizot2006}.
By varying the confinement the crossover from three-dimensional~\cite{greiner2002,duan2003} to two-dimensional~\cite{zobay2001,colombe2004} to one-dimensional~\cite{orzel2001,paredes2004} traps can be tuned.
Feshbach~\cite{kohler2006,chin2010} and confinement-induced resonances~\cite{olshanii1998,kim2006,giannakeas2012,giannakeas2013} offer fine-grained control of the inter-particle interaction.
Recent studies within the realm of ultracold atoms provide close links to solid-state systems~\cite{anderson1998,jo2009}, electronic structure of molecules~\cite{luhmann2015}, light-matter interaction~\cite{sala2017}, topological matter~\cite{jotzu2014,goldman2016} and black-hole analogs~\cite{steinhauer2016}.
The increasing progress of the experimental control of these many-body systems demands appropriate theoretical and numerical methods to describe them and to calculate their properties as well as their dynamical behavior.
Exactly solvable models are rare while usually relying on more or less crude approximations or focusing on certain limiting cases.

Let us discuss the state of the art of analytically solvable models and numerical approaches.
The time-dependent Schrödinger equation of two bosons in a parabolic and spherically symmetric trapping potential is exactly solvable~\cite{busch1998}.
However, the applicability of such a small system is very limited.
Larger particle numbers can be studied using the Lieb-Liniger model~\cite{lieb1963,lieb1963a} for spinless bosons with contact interactions~\cite{pethick2008} assuming periodic boundary conditions.
Yet, this approach is not capable of taking external trapping potentials into account and cannot directly describe the dynamical response of the system.
The Tonks-Girardeau~\cite{girardeau1960,yukalov2005} model on the other hand grants access to the full many-body spectrum and nonequilibrium solutions by mapping bosons to non-interacting fermions.
However, this model is only valid in the limit of inifinitely strong interactions and in one spatial dimension.
Beyond these limitations of analytical approaches powerful computational methods are needed to study ensembles of ultracold atoms.

A very useful approach is the \gls{gpe}~\cite{gross1961,pitaevskii1961} which represents a non-linear Schrödinger equation for a bosonic many-body ensemble in the presence of an external trap with contact inter-particle interaction in the thermodynamic limit.
It assumes the Hartree-Fock approximation~\cite{hartree1928,fock1930} to the many-body wave function, leading to an effective, mean-field description.
The \gls{gpe} is a partial differential equation which can be solved efficiently using the typical finite element and finite difference methods~\cite{cerimele2000,bao2003,dion2003,muruganandam2003,muruganandam2009,danaila2010,danaila2010a}.
This mean-field treatment allows for the study of setups containing large particle numbers and enables the description of a multitude of non-linear wave structures such as dark and bright solitons~\cite{burger1999,alkhawaja2002}.
In some cases, when potential and interaction energy dominate the kinetic energy, the calculation can be further simplified by ignoring the kinetic term of the Schrödinger equation leading to the Thomas-Fermi~\cite{baym1996} approximation.
In general however, these mean-field descriptions do not provide an adequate description of the system dynamics as they cannot account for quantum correlations.
A prominent example where the \gls{gpe} fails to capture the correct physical behavior is the bosonic Josephson junction~\cite{sakmann2009,sakmann2014}.
For weakly depleted condensates Bogoliubov theory~\cite{bogoliubov1946,lee1957a,lee1957} can be applied.
For the investigation of few- to many-body systems with substantial correlations and correlated dynamics, however, ab initio beyond-mean-field methods are necessary.

One of the most fundamental of such methods is the exact diagonalization treatment of the many-body Hamiltonian~\cite{damski2005,zhang2010,hu2012} which grants access to the spectrum and the eigenstates of the physical system.
However, this approach is limited to a small number of particles due to the computational complexity of diagonalization algorithms.
Furthermore, the choice of an appropriate basis can prove difficult so that a large number of basis functions may be required, thereby further enlarging the numerical effort.
This computational challenge calls for more efficient numerical approaches.

Many computational approaches focus on the investigation of optical lattices as these setups are of major interest in the research of ultracold neutral atoms due to the condensed matter counterparts (crystals).
Often the Bose-Hubbard model~\cite{jaksch1998,fisher1989} is employed to describe bosonic atoms loaded into the lowest band of a sufficiently deep lattice.
In this model, the bosonic field operator is expanded into Wannier states yielding an effective theoretical model, where the kinetic term as well as the trapping potential are reduced to a hopping between lattice sites and the interaction term to an on-site interaction.
This model has been studied using a plethora of different methods~\cite{lewenstein2007} including \gls{dmrg}~\cite{kuhner1998,rapsch1999} and \gls{qmc}~\cite{purwanto2004,wessel2004}.
However, other approaches are required to describe physical systems and effects beyond the applicability of a Hubbard model, covering in particular their out-of-equilibrium dynamics.

The \gls{mctdh}~\cite{meyer1990,beck2000} is such a method and has proven to be a powerful and versatile tool to ab initio solve the time-dependent Schrödinger equation for correlated many-body systems of distinguishable degrees of freedom ab initio.
\gls{mctdh} has been extended to study fermionic ensembles using the \gls{mctdhf} method~\cite{zanghellini2003,caillat2005} and for bosonic systems using the \gls{mctdhb}~\cite{streltsov2007,alon2008} rendering the treatment of ultracold atoms possible.
Further extensions~\cite{kronke2013,cao2013,cao2017} employing a multi-layer approach also allow for the treatment of Bose-Bose mixtures and more recently Bose-Fermi and Fermi-Fermi mixtures further increasing the usefulness and applicability of this family of methods.
The power of this class of methods stems from the usage of a variationally optimized, time-dependent set of basis functions that allows for a compact representation of the many-body wave function and yields a beyond-mean-field description that takes all correlations into account.

As all numerical approaches however, \gls{mctdhb} faces the problem of an exponentially growing Hilbert space when studying large many-body systems.
In particular, when increasing either the number of particles or the size of the \gls{spf} basis used to describe such an atomic ensemble, the number of possible number states or configurations respectively grows rapidly rendering the treatment of systems typically with particle numbers larger than one hundred (in ther superfluid regime) challenging if not computationally prohibitive.
To tackle this issue within the family of \gls{mctdh} methods, different approaches have been proposed in the literature.
For instance, the configuration selection schemes for \gls{mctdh}~\cite{worth2000,wodraszka2016,wodraszka2017} or the \gls{ras} schemes for \gls{mctdhf}~\cite{miyagi2013,miyagi2014} and \gls{mctdhb}~\cite{leveque2017} perform a static selection of the most relevant Hartree products/Slater determinants for the physical system and exploit this partitioning to reduce the required numerical effort.
However, these methods cannot dynamically adapt to the evolution of the system and require a priori knowledge such as the choice of an excitation scheme in the case of \gls{ras}.
Such static truncation schemes of the Hilbert space can impose artificial constraints on the physical system if important many-body states are removed.
Therefore the development of dynamical, self-adapting approaches is required in order to enable a more general treatment of dynamical many-body systems.

Referring to the investigation of distinguishable degrees of freedom, dynamical procedures have been applied successfully within the framework of \gls{mctdh}.
E.g.\ by pruning the primitive basis/grid~\cite{larsson2016,larsson2017} or the coefficients of the wave function~\cite{larsson2017} the runtime of the simulations can be greatly reduced.
Unfortunately however, the pruning of the grid is not very lucrative in calculations with ultracold atoms as these ensembles are usually confined using an external potential so that there rarely exist unoccupied regions of real space.
Furthermore, the coefficient based pruning approach presented in Ref.~\cite{larsson2017} cannot be applied as the proposed neighborship criterion for the coefficients cannot be easily transferred to the number states of indistinguishable particles.
Therefore the development of new dynamical methods for the treatment of indistinguishable particles is necessary.

In the present work, we develop a general method that automatically detects the important number states of bosonic many-body systems when studying the nonequilibrium dynamics using \gls{mctdhb}.
This selection procedure dynamically adapts during the time evolution of the system.
In Section~\ref{sec:mctdhb} we start by briefly reviewing the \gls{mctdhb} theory in order to introduce the key concepts of this method and motivate our pruning algorithm.
In Section~\ref{sec:pruning} we show two different ways to modify the \gls{mctdhb} \glspl{eom} in order to reduce the numerical effort.
To achieve this, we introduce a pruning threshold and a selection criterion for determining the importance of each number state.
In Section~\ref{sec:criteria} we present two different criteria for the selection of the number states relying on the overlap with the many-body wave function and the total energy of the system.
To showcase the usefulness of our approach, we benchmark it using two different physical scenarios in Section~\ref{sec:applications}.
We focus both on the performance benefits as well as the accuracy when compared to a regular \gls{mctdhb} simulation.
Finally, we summarize our findings in Section~\ref{sec:conclusions} and discuss future perspectives of our approach.
In Appendix~\ref{sec:convergence} we comment on the convergence of our numerical results.

\section{\label{sec:mctdhb}Key aspects of the \glsentrylong{mctdhb}}
\gls{mctdhb} allows to describe the correlated quantum dynamics of ensemble of $N$ interacting bosons.
It employs a variationally optimal, time-dependent basis $\lbrace\varphi_i(t)\rbrace_{i=1}^{m}$ of $m$ \glsentryfullpl{spf} also called orbitals.
Compared to other methods that employ a stationary basis, significantly fewer basis functions are required to achieve the same level of description of correlations.
The many-body wave function is expanded as a superposition
\begin{equation}
  \ket{\Psi(t)}=\sum\limits_{\vec{n}\in\mathcal{V}}C_{\vec{n}}(t)\ket{\vec{n};t}
\end{equation}
of all $N_{\mathcal{V}}=\binom{N+m-1}{N}$ time-dependent permanents ${\lbrace\ket{\vec{n};t}\rbrace}_{\vec{n}\in\mathcal{V}}$ that retain the total number of particles $N$ using time-dependent coefficients ${\lbrace C_{\vec{n}}(t)\rbrace}_{\vec{n}\in\mathcal{V}}$.

Each vector $\vec{n}={\left(n_1\; n_2\; \cdots\; n_m\right)}^{\mathrm{T}}$ resembles one way of distributing $N$ particles in $m$ orbitals and is called a configuration.
The $i$th component $n_i$ of such a vector specifies the number of particles in the orbital $\varphi_i(t)$ for the given configuration.
$\mathcal{V}=\lbrace\vec{n}\in\mathbb{N}_0^m: {\|\vec{n}\|}_1=N\rbrace$ is the set of all such configurations that with the total number of particles $N$.

The permanents are given by
\begin{equation}
  \ket{\vec{n};t}=\left(\prod\limits_{i=1}^m\frac{{\left(a_i^\dagger(t)\right)}^{n_i}}{\sqrt{n_i!}}\right)\ket{\vec{0}}
\end{equation}
in terms of the bosonic creation operators ${\lbrace a_i^\dagger(t)\rbrace}_{i=1}^m$ with respect to the instantaneous basis.

\gls{mctdhb} solves the time-dependent Schrödinger equation $\left(i\hbar\partial_t -\hat{H}(t)\right)\ket{\Psi(t)}=0$ as an initial value problem by propagating an initial wave function $\ket{\Psi(0)}$ in time according to a, potentially time-dependent, Hamilton operator $\hat{H}(t)$.
In this work we limit ourselves to Hamiltonians of the form
\begin{equation}
  \hat{H}(t,\lbrace x_i\rbrace)=\sum\limits_{i=1}^{N}\hat{h}(t,x_i)+\sum\limits_{i<j}\hat{W}(t,x_i,x_j),
\end{equation}
containing only one-body ($\hat{h}$) and two-body ($\hat{W}$) terms.
By employing the Lagrangian~\cite{broeckhove1988}, Dirac-Frenkel~\cite{dirac1930,frenkel1932a} or McLachlan~\cite{mclachlan1964} variational principle, one can derive the corresponding \gls{mctdhb} \glspl{eom}~\cite{alon2008,alon2007,streltsov2007} which are integrodifferential equations describing the time evolution of the coefficients ${\lbrace C_{\vec{n}}(t)\rbrace}_{\vec{n}\in\mathcal{V}}$ and the \glspl{spf} ${\lbrace\varphi_i(t)\rbrace}_{i=1}^m$.

The \gls{spf} \gls{eom} describes a rotation of the orbitals in such a way that they represent the state of the physical system optimally.
For details on this equation we refer the reader to Ref.~\cite{alon2008} as the precise structure is irrelevant for the pruning approach that we describe herein.
The time evolution of the time-dependent coefficients ${\lbrace C_{\vec{n}}(t)\rbrace}_{\vec{n}\in\mathcal{V}}$ is governed by
\begin{equation}
  i\partial_t C_{\vec{n}}(t)=\sum\limits_{\vec{m}\in\mathcal{V}}\underbrace{\bra{\vec{n};t}\hat{H}(t)\ket{\vec{m};t}}_{H_{\vec{n}\vec{m}}(t)}C_{\vec{m}}(t)\label{eq:eom_coeff}
\end{equation}
which is coupled to the \glspl{eom} of the orbitals via the configurations $\ket{\vec{n};t}$.

\section{\label{sec:pruning}Pruned Equations of Motion}
The number of possible configurations $N_{\mathcal{V}}$ grows rapidly with the number of particles $N$ and orbitals $m$.
This scaling behavior renders the treatment of large systems challenging if not infeasible as the number of matrix elements grows quadratically with $N_{\mathcal{V}}$ causing the integration of Equations~\eqref{eq:eom_coeff} to become very costly and the dominant contribution to the simulation runtime.

From intuition and experience we know that not all configurations are of equal importance for the corresponding physical systems under consideration.
In the present work, we establish measures to automatically detect configurations of lesser importance and leverage this knowledge to reduce the numerical effort of the integration of the coefficient \glspl{eom}.
Our approach is dynamical and regularly reevaluates the importance of all configurations, in particular also those that have been deemed negligible previously.

In order to derive our pruning approach, we start by defining a measure $f:\mathcal{V}\times\mathbb{R}\to\mathbb{R}$ that determines the importance of each configuration $\vec{n}$ at time $t$.
We divide the set $\mathcal{V}$ of all configurations into the subset of unpruned (i.e.\ active) configurations
\begin{equation}
  \mathcal{P}(t)=\lbrace\vec{n}:f(\vec{n},t)>\gamma\rbrace
\end{equation}
and the subset of pruned (i.e.\ inactive) configurations
\begin{equation}
  \label{eq:Q}
  \mathcal{Q}(t)=\lbrace\vec{n}:f(\vec{n},t)\leq\gamma\rbrace
\end{equation}
by introducing a pruning threshold $\gamma\in\mathbb{R}$.
Additionally, we introduce the operators
\begin{align}
  \hat{P}(t)=\sum_{\vec{m}\in\mathcal{P}(t)}\dyad{\vec{m};t}{\vec{m};t} \\
  \hat{Q}(t)=\sum_{\vec{m}\in\mathcal{Q}(t)}\dyad{\vec{m};t}{\vec{m};t}
\end{align}
that project onto the configuration subsets $\mathcal{P}(t)$ and $\mathcal{Q}(t)$.
In the following, we drop the explicit notation of the time-dependence of the sets and the projection operators for the sake of readability.
The many-body Hamiltonian can be rewritten in terms of $\hat{P}$ and $\hat{Q}$ as
\begin{equation}
  \label{eq:hamiltonian_projectors}
  \begin{split}
      \hat{H}&=\hat{P}\hat{H}\hat{P}+\hat{Q}\hat{H}\hat{P}+\hat{P}\hat{H}\hat{Q}+\hat{Q}\hat{H}\hat{Q} \\
             &=\hat{H}\hat{P}+\hat{P}\hat{H}\hat{Q}+\hat{Q}\hat{H}\hat{Q},
  \end{split}
\end{equation}
where we exploited the property $\hat{P}+\hat{Q}=\hat{1}$.

\begin{figure}[ht]
  \includegraphics[width=0.45\textwidth]{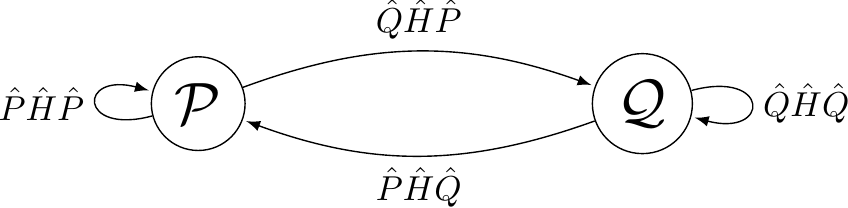}
  \caption{\label{fig:terms} Illustration of the meaning of the different terms of $\hat{H}$ with respect to projection operators.}
\end{figure}

The idea of the pruning approach is to neglect terms in this representation, thus defining a new, truncated Hamiltonian which replaces the original in Equation~\eqref{eq:eom_coeff}.
In order to make an adequate choice for the pruning, it is essential to consider the meaning of each term of Eq.~\eqref{eq:hamiltonian_projectors} within the context of the coefficient \glspl{eom}~\eqref{eq:eom_coeff}, see also Figure~\ref{fig:terms}.
The term $\hat{Q}\hat{H}\hat{Q}$ mediates between number states belonging to configurations from $\mathcal{Q}$, i.e.\ configurations that we consider negligible.
Therefore, the most apparent modification is to neglect this part of the Hamiltonian yielding
\begin{equation}
  \hat{H}_1^\prime=\hat{H}\hat{P}+\hat{P}\hat{H}\hat{Q}\label{eq:hp_phq}.
\end{equation}
We note that $H_1^\prime$ is Hermitian.
When inserting this Hamiltonian into Equation~\eqref{eq:eom_coeff}, the resulting modified \gls{eom} reads
\begin{equation}
  \label{eq:eom_hp_phq}
  \begin{split}
    i\partial_t C_{\vec{n}}(t)=&\sum\limits_{\vec{m}\in\mathcal{P}(t)}\bra{\vec{n};t}\hat{H}(t)\ket{\vec{m};t}C_{\vec{m}}(t)\\
    +&\underbrace{\sum\limits_{\vec{m}\in\mathcal{Q}(t)}\bra{\vec{n};t}\hat{H}(t)\ket{\vec{m};t}C_{\vec{m}}(t)}_{\text{if }\vec{n}\in\mathcal{P}(t)},
  \end{split}
\end{equation}
where the second term is only present for coefficients associated with configurations from $\mathcal{P}$.
However, we note that only the right-hand side of the \glspl{eom} is modified while the total number of \glspl{eom} remains unchanged such that all coefficients ${\lbrace C_{\vec{n}}(t) \rbrace}_{\vec{n}\in\mathcal{V}}$ are propagated in time.
This is a key element of our approach as it allows coefficients corresponding to inactive configurations (from the set $\mathcal{Q}$) to evolve such that they could be activated again should they transcend the pruning threshold.

In the present work, we also investigate a second type of pruned Hamiltonian
\begin{equation}
  H_2^\prime =\hat{H}\hat{P}\label{eq:hp}
\end{equation}
that we obtain by also neglecting the term $\hat{P}\hat{H}\hat{Q}$ which mediates scattering from the negligible configurations $\mathcal{Q}$ to the active configurations $\mathcal{P}$.
However, this operator is non-Hermitian but our numerical results in Sec.~\ref{sec:applications} suggest that this weak non-hermiticity may still be acceptable in the sense that the many-body dynamics can still be described to some accuracy (see Section~\ref{sec:applications}).
The corresponding \gls{eom} reads
\begin{equation}
  \label{eq:eom_hp}
  i\partial_t C_{\vec{n}}(t)=\sum\limits_{\vec{m}\in\mathcal{P}(t)}\bra{\vec{n};t}\hat{H}(t)\ket{\vec{m};t}C_{\vec{m}}(t).
\end{equation}

In the standard \gls{mctdhb} algorithm, an initial wave function $\ket{\Psi(t_0)}$ is propagated from the initial time $t_0$ to a final time $t_{\mathrm{f}}$ using some time step $\Delta t$ at which the wave function is to be computed.
The interval $\Delta t$ is usually divided further due to the usage of an adaptive integrator~\cite{dormand1980,press2007a,stoer2002a}.
Algorithm~\ref{alg:pruning} shows how we integrate the pruning approach into this existing procedure.
We introduce an additional timescale $\tau$ that determines when the pruning criterion is to be evaluated.
The resulting selection of active configurations is kept constant for the time $\tau$.
Initially, all configurations are marked as active as can be seen in Line~\ref{alg:line:init_P} and~\ref{alg:line:init_Q}.
The initial wave function is then propagated until the target time $t_{\mathrm{f}}$ is reached (see Lines~\ref{alg:line:begin_while}--\ref{alg:line:end_while}).
Whenever the time $\tau$ has passed, the pruning criterion is reevaluated and the selection of active configurations is updated (see Lines~\ref{alg:line:begin_pruning}--\ref{alg:line:end_pruning}).
\begin{algorithm}[H]
  \caption{\label{alg:pruning} Propagation procedure for pruned \mbox{\gls{mctdhb}} simulations}
  \begin{algorithmic}[1]
    \Procedure{PrunedPropagation}{$\ket{\Psi(t_0)}$, $t_0$, $t_{\mathrm{f}}$, $\Delta t$, $\tau$, $\gamma$}
      \State $t_{\text{next}}\gets t_0+\Delta t$
      \State $t_{\text{pruning}}\gets t_0+\tau$
      \State $\mathcal{P}\gets\mathcal{V}$\label{alg:line:init_P}
      \State $\mathcal{Q}\gets\lbrace\rbrace$\label{alg:line:init_Q}
      \While{$t<t_{\mathrm{f}}$}\label{alg:line:begin_while}
        \State $t^\prime\gets\min\lbrace t_{\text{next}}, t_{\text{pruning}}\rbrace$
        \State $\ket{\psi(t^\prime)} \gets \text{propagate}(t,t^\prime,\ket{\psi(t)},\mathcal{P},\mathcal{Q})$\label{alg:line:propagate}
        \If{$t^\prime = t_{\text{pruning}}$}\label{alg:line:begin_pruning}
          \State $\mathcal{P}\gets\lbrace\vec{n}:f(\vec{n},t)>\gamma\rbrace$
          \State $\mathcal{Q}\gets\lbrace\vec{n}:f(\vec{n},t)\le\gamma\rbrace$
          \State $t_{\text{pruning}}\gets t_{\text{pruning}}+\tau$
        \EndIf\label{alg:line:end_pruning}
        \If{$t^\prime = t_{\text{next}}$}
          \State write $\ket{\Psi(t)}$, evaluate observables, etc.
          \State $t_{\text{next}}\gets t_{\text{next}}+\Delta t$
        \EndIf
        \State $t\gets t^\prime$
      \EndWhile\label{alg:line:end_while}
    \EndProcedure
  \end{algorithmic}
\end{algorithm}

Both the pruning time $\tau$ and the threshold $\gamma$ impact the dynamical pruning algorithm.
Choosing small values of $\gamma$ reduces the ratio of configurations that can be disabled on the right hand side of each \gls{eom} (see Equation~\eqref{eq:eom_hp_phq} and~\eqref{eq:eom_hp}) and thus the speedup that can be achieved.
In the case $\gamma=0$, all configurations are taken into account and the dynamical pruning approach is equivalent to the original \gls{mctdhb}.
However, choosing $\gamma$ very large may lead to incorrect results as important number states might be neglected.
As the pruning time $\tau$ determines how often the pruning criterion is evaluated, this parameter has to be chosen appropriately depending on the timescales of the physical system.
Small values of $\tau$ lead to very frequent reevaluations of the pruning criterion which can negate any performance gain due to the decreased number of configurations as the evaluation of the criterion introduces additional computational effort.
When using large values for $\tau$ the evolution of the physical system might be imprecise.
To ensure that the numerical results are accurate enough $\tau$ and $\gamma$ have to be chosen carefully by learning how to handle them via the comparison with converged results, e.g.\ the original \gls{mctdhb} results.

\section{\label{sec:criteria}Pruning Criteria}
In Section~\ref{sec:pruning} we outlined our pruning approach and introduced the function $f(\vec{n},t)$ without further specifying it.
In the following, we present two different pruning criteria that we use for the applications in Section~\ref{sec:applications}.
We base our choices on the norm of the wave function and the total energy as these quantities are easily accessible and interpretable.

\subsection{\label{sec:magnitude_criterion}Magnitude Criterion}
The most obvious way to assess the importance of a configuration is to project the many-body wave function onto the corresponding number state and compute the magnitude of the overlap
\begin{equation}
  \label{eq:mc}
  f(\vec{n},t)={\left|\ip{\Psi(t)}{\vec{n};t}\right|}^2={\left|C_{\vec{n}}(t)\right|}^2.
\end{equation}
This criterion, which we refer to as the \gls{mc} in the following, is intuitive as we can compute a real number $f(\vec{n},t)\in\left[0,1\right]$ that determines the importance of the configuration $\vec{n}$.
A value of $0$ means that the configuration does not contribute at all to the many-body wave function whereas a value of $1$ implies that the wave function is given solely by the corresponding number state.

\subsection{\label{sec:energy_criterion}Energy Criterion}
In order to investigate the impact of the pruning criterion on the numerical results, we study a second possible choice.
For the so-called \gls{ec} we determine the contribution of a configuration to the total energy.
The energy of a \gls{mctdhb} wave function is given by
\begin{equation}
  \label{eq:energy}
  E(t)=E_{\mathcal{V}}(t)=\sum\limits_{\vec{n},\vec{m}\in\mathcal{V}}C_{\vec{n}}^*(t)C_{\vec{m}}(t)H_{\vec{n},\vec{m}}(t)
\end{equation}
and depends solely on the time-dependent coefficients $\lbrace C_{\vec{n}}(t)\rbrace$.
In order to estimate the energetic contribution of a single, specific configuration $\vec{n}$ we expand $E_{\mathcal{V}}(t)$ as a Taylor polynomial of first order with respect to the corresponding coefficient
\begin{equation}
  E_{\mathcal{V}}(t)\approx E_{\mathcal{V}\backslash\lbrace\vec{n}\rbrace}(t)+\frac{\partial E_{\mathcal{V}}(t)}{\partial C_{\vec{n}}^*(t)}C_{\vec{n}}^*(t),
\end{equation}
where $E_{\mathcal{V}\backslash\lbrace\vec{n}\rbrace}(t)$ is the energy of the system when neglecting the configuration $\vec{n}$, i.e.\ setting $C_{\vec{n}}(t)$ to $0$.
Consequently, an estimate of the absolute, energetic contribution of the configuration $\vec{n}$ is given by
\begin{equation}
  E_{\vec{n}}(t)=E_{\mathcal{V}}(t)-E_{\mathcal{V}\backslash\lbrace\vec{n}\rbrace}(t)=\frac{\partial E_{\mathcal{V}}(t)}{\partial C_{\vec{n}}^*(t)}C_{\vec{n}}^*(t).
\end{equation}
We normalize this quantity by dividing by the total energy and taking the absolute value
\begin{equation}
  f(\vec{n},t)=\Delta E_{\vec{n}}(t)=\left|\frac{E_{\vec{n}}(t)}{E_{\mathcal{V}}(t)}\right|=\left|\frac{1}{E_{\mathcal{V}}(t)}\frac{\partial E_{\mathcal{V}}(t)}{\partial C_{\vec{n}}^*(t)}C_{\vec{n}}^*(t)\right|\label{eq:ec}
\end{equation}
in order to obtain a real number $f(\vec{n},t)\in\left[0, 1\right]$ which can be interpreted as the relative energy contribution.

\section{\label{sec:applications}Application to the Quantum Dynamics of Trapped Ultracold Bosonic Ensembles}
In the following, we consider a one-dimensional system of $N$ identical bosons confined in an external potential $V(x)$.
Note that within the ultracold regime $s$-wave scattering is the dominant interaction process~\cite{pitaevskii2003,pethick2008} such that we consider contact interactions between the particles.
The many-body Hamiltonian of such a system is given by
\begin{equation}
  \begin{split}
    \hat{H}\left(\lbrace x_i \rbrace\right)=&\sum\limits_{i=1}^N \left(-\frac{\hbar²}{2m}\frac{\partial^2}{\partial x_i^2}+V(x_i)\right) \\
    +&\sum\limits_{i<j}g\delta(x_i - x_j).
  \end{split}
\end{equation}

Starting from the ground state of the non-interacting system (i.e. $g=0$) we compute the many-body ground state of the interacting ensemble by imaginary time propagation leading to energy relaxation~\cite{kosloff1986,beck2000} or via the improved relaxation algorithm~\cite{meyer2003}.
The resulting initial ground state wave function is then propagated in time with respect to a quenched Hamiltonian which involves an instantaneous change in one of the system parameters~\cite{greiner2002,greiner2002b,polkovnikov2011}.
The propagation of the wave function is performed using the usual, unpruned \gls{mctdhb} as well as the pruned variants introduced in Sec.~\ref{sec:pruning}.
To determine the benefits of pruning we measure the CPU time of all simulations and also monitor the number of configurations that are pruned at each time step.
In order to quantify the amount of deactivated configurations we define the quantity
\begin{equation}
  \label{eq:pruning_ratio}
  \beta(t)=\frac{\left|\mathcal{Q}(t)\right|}{\left|\mathcal{V}(t)\right|}
\end{equation}
as the ratio between the number of inactive configurations (the cardinality of the set $\mathcal{Q}(t)$, see Equation~\eqref{eq:Q}) and the total number of configurations (the cardinality of the set $\mathcal{V}$, see Section~\ref{sec:mctdhb}).
Additionally, we compare different physical quantities between the pruned and unpruned \gls{mctdhb} data in order to assess the accuracy of our pruning approach.

\gls{mctdhb} provides the full many-body wave function at each time step of the evolution and thus grants access to a plethora of different observables that allow us to analyze and understand the physical system.
One of the most general of such quantities is the reduced $p$-body density matrix~\cite{sakmann2008}
\begin{widetext}
\begin{equation}
  \rho_p(x_1,\ldots,x_p,x_1^\prime,\ldots,x_p^\prime,t)=\frac{N!}{(N-p)!}\int\Psi(x_1,\ldots,x_N,t)\Psi^*(x_1^\prime,\ldots,x_p^\prime,x_{p+1},\ldots,x_N,t)\diff{x_{p+1}}\ldots\diff{x_N}
\end{equation}
\end{widetext}
that can be used to calculate particle densities as well as correlation functions.
The reduced one-body density matrix (i.e.\ $p=1$) is of special interest as its diagonal $\rho_1(x,t)=\rho_1(x,x^\prime=x,t)$ is the one-body density which describes the spatial distribution of particles.
The spectral representation of the reduced one-body density matrix
\begin{equation}
  \rho_1(x,x^\prime,t)=\sum\limits_{\alpha=1}^{m} \lambda_{\alpha}(t)\phi_{\alpha}(x,t)\phi_{\alpha}^*(x^\prime,t)
\end{equation}
is given by the eigenvectors $\lbrace\phi_{\alpha}(t)\rbrace$, the so-called natural orbitals, and the decreasingly ordered eigenvalues $\lambda_{\alpha}(t)\in\left[0,1\right]$, the so-called natural populations.
The natural populations fulfill $\sum_{\alpha=1}^m \lambda_{\alpha}(t)=1$ and determine the degree of inter-particle correlations within the ensemble.
A system with $\lambda_{1}=1\wedge\lambda_{\alpha > 1}=0$ is called condensed, is accurately described in a mean-field treatment using a single orbital and does not exhibit inter-particle correlations.
In order to quantify the impact of the pruning approach on the natural populations, we compute the absolute difference between the natural population $\lambda_i(t)$ obtained by a regular \gls{mctdhb} calculation and the corresponding natural population $\lambda_i^\prime(t)$ from a pruned simulation
\begin{equation}
  \label{eq:absolute_natpop_error}
  \varepsilon_{\lambda_i}(t)=\left|\lambda_i^\prime(t)-\lambda_i(t)\right|.
\end{equation}

Moreover, we study the reduced two-body density operator which can be used to calculate second order correlation functions and two-particle densities.
For the sake of brevity we only report results on the diagonal $\rho_2(x, x)=\rho_2(x_1=x,x_2=x,x_1^\prime=x,x_2^\prime=x,t)$ that can be interpreted as the probability distribution to find two particles at the same position.

Furthermore, the pruning might affect the total energy of the system due to the modifications of the many-body Hamiltonian.
In order to quantify any such effects, we introduce the relative deviation
\begin{equation}
  \label{eq:relative_energy_error}
  \varepsilon_{E}(t)=\left|\frac{E^\prime(t)}{E(t)}-1\right|.
\end{equation}
between the energy $E^\prime(t)$ of a pruned calculation and $E(t)$ of a regular \gls{mctdhb} calculation.

The \gls{mctdhb} algorithm conserves the norm of the wave function, the orthonormality of the \glspl{spf} and, if the Hamiltonian is time-independent, the energy.
The dynamical pruning might introduce inaccuracies that lead to violations of theses properties which we investigate by defining appropriate error quantities.
The norm of the wave function should always have the value $1$.
By computing the absolute difference
\begin{equation}
  \label{eq:norm_conservation}
  \xi_{{\left\|\Psi\right\|}^2}(t)=\left|{\ip{\Psi(t)}{\Psi(t)}}^2-1\right|
\end{equation}
from this target value we can quantify the violation of the norm conservation at each time $t$.
In Sections~\ref{sec:lattice} and~\ref{sec:ho} we study the dynamics after a sudden change of the Hamiltonian at time $t=0$.
However, we keep the Hamiltonian $H(t\ge 0)$ constant such that the energy should be conserved throughout the simulation.
In order to measure violations of this conservation law, we compute the relative difference
\begin{equation}
  \label{eq:energy_conservation}
  \xi_{E}(t)=\left|\frac{E(t)}{E(0)}-1\right|
\end{equation}
of the momentary total energy $E(t)$ with respect to the initial energy $E(0)$.
During the propagation of a many-body wave function using \gls{mctdhb} the \gls{spf} basis should remain orthonormal.
To quantify deviations from this property at time $t$ we introduce the quantity
\begin{equation}
  \xi_{\perp}(t)=\ip{\varphi_i(t)}{\varphi_j(t)}-\delta_{ij}.
\end{equation}
However, we do not discuss this property in detail in the following sections as we find that it does not seem to be affected by the pruning approach.
The value of this quantity is always comparable to the unpruned \gls{mctdhb} calculation and is bounded by $\xi_{\perp}(t)<10^{-10}$.

In the following, we discuss two physical scenarios by choosing different potentials and quench procedures to showcase the performance of the pruning approach.
To ensure comparability, all numerical simulations where performed on an AMD\textsuperscript{®} Ryzen™ Threadripper™ 1950X 16-core processor using a single, dedicated core.

\subsection{\label{sec:lattice}Quench Dynamics in an Optical Lattice}
We investigate the nonequilibrium dynamics of repulsively interacting bosons trapped in an optical lattice~\cite{jaksch1998} following a sudden change of the interaction strength.
This quench procedure is experimentally accessible through Feshbach resonances~\cite{kohler2006,chin2010} or changes of the transversal confinement frequency~\cite{kim2006,giannakeas2012,giannakeas2013}.
Similar setups have been investigated using \gls{mctdhb} in several previous works~\cite{mistakidis2014,lode2015,mistakidis2015,koutentakis2017,mistakidis2017,neuhaus-steinmetz2017,mistakidis2018,plassmann2018} so that this setup serves as an ideal testbed for new methodological advancements.

We parameterize the lattice potential as
\begin{equation}
  V(x)=\begin{cases}
    V_0\sin^2\left(\frac{\pi p x}{L}\right) & -\frac{L}{2}\leq x \leq\frac{L}{2} \\
    \infty & \text{otherwise}
  \end{cases}
\end{equation}
with an odd number of wells $p$, the barrier height $V_0$ and the system size $L$.
Based on these lattice parameters we use the recoil energy~\cite{jaksch2005} $E_{\mathrm{R}}=\hbar^2\pi^2 p^2/2mL^2$ as the natural energy unit of the system and choose a barrier height of $V_0/E_{\mathrm{R}}=4$ for the lattice in the following.
Starting from the ground state of the system with $\tilde{g}=p \pi g/L E_{\mathrm{R}}=0.1$, we study the dynamics following a quench to $\tilde{g}=0.4$ and $\tilde{g}=0.8$.
According to the convergence checks that we performed (see Appendix~\ref{sec:convergence}) it proves sufficient to restrict the \gls{spf} basis to $m=5$ orbitals.

\begin{figure}[ht]
  \centering
  \includegraphics[width=0.45\textwidth]{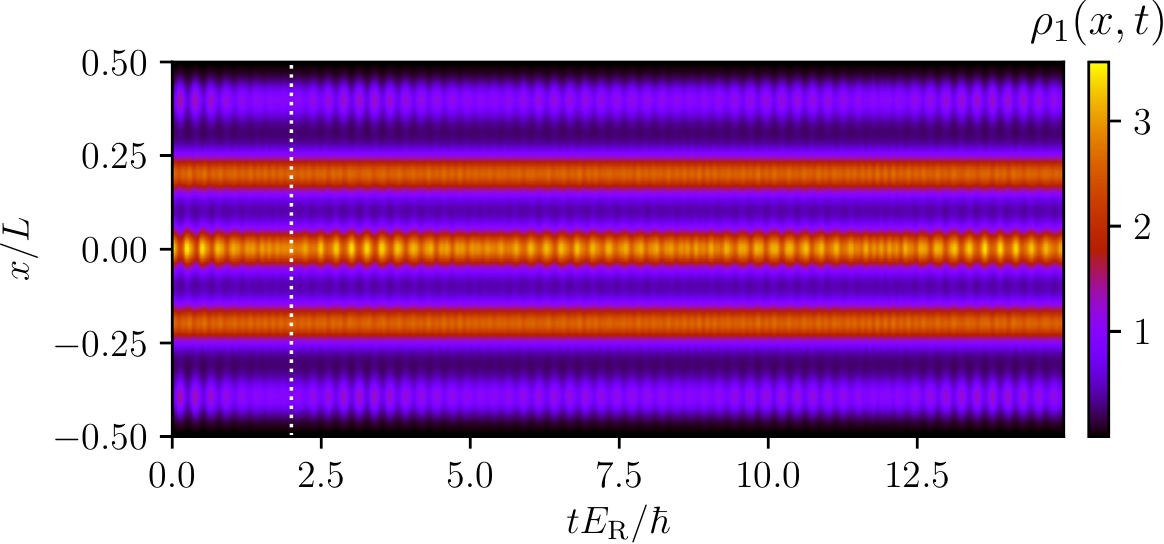}
  \caption{\label{fig:lattice_gpop} Time evolution of the one-body density $\rho_1(x,t)$ for $N=20$ bosons in a finite five-well lattice following an interaction quench from $\tilde{g}=0.1$ to $\tilde{g}=0.4$ according to a regular \gls{mctdhb} calculation. The dotted white line indicates the final propagation time $t_{\mathrm{f}}=2\hbar/E_{\mathrm{R}}$ that is used for the comparison with the different pruning approaches for different numbers of particles and post-quench interaction strengths.}
\end{figure}

Figure~\ref{fig:lattice_gpop} shows the evolution of the one-body density after the aforementioned quench protocol for the case of $N=20$ particles and a quench to $\tilde{g}=0.4$ which has been computed using a regular \gls{mctdhb} simulation.
The quench excites intrawell breathing dynamics which is visible as a periodic expansion and contraction of the atomic cloud around the center of each well.
Additionally, over-barrier transport between the wells~\cite{mistakidis2014} is induced which can be identified by the finite particle density between the wells.
In Figure~\ref{fig:lattice_gpop} we show the propagation up to a final time $t_{\mathrm{f}}\approx 15\,\hbar/E_{\mathrm{R}}$.
In the further analysis we investigate the different pruning approaches for a varying number of particles and post-quench interaction strengths leading to a large number of independent simulations.
As the simulation times can become large, especially when treating larger numbers of particles, we simplify the analysis by only studying the dynamics up to a final time $t_{\mathrm{f}}=2\,\hbar/E_{\mathrm{R}}$ as indicated by the dotted white line in Figure~\ref{fig:lattice_gpop}.
In order to ensure that our pruning approach also captures the correct long-term behavior of the physical system, we also performed calculations up to a time $t=10\hbar/E_{\mathrm{R}}$ for a selection of these simulations, the results of which we do not present for the sake of brevity.

\begin{figure}[ht]
  \centering
  \includegraphics[width=0.45\textwidth]{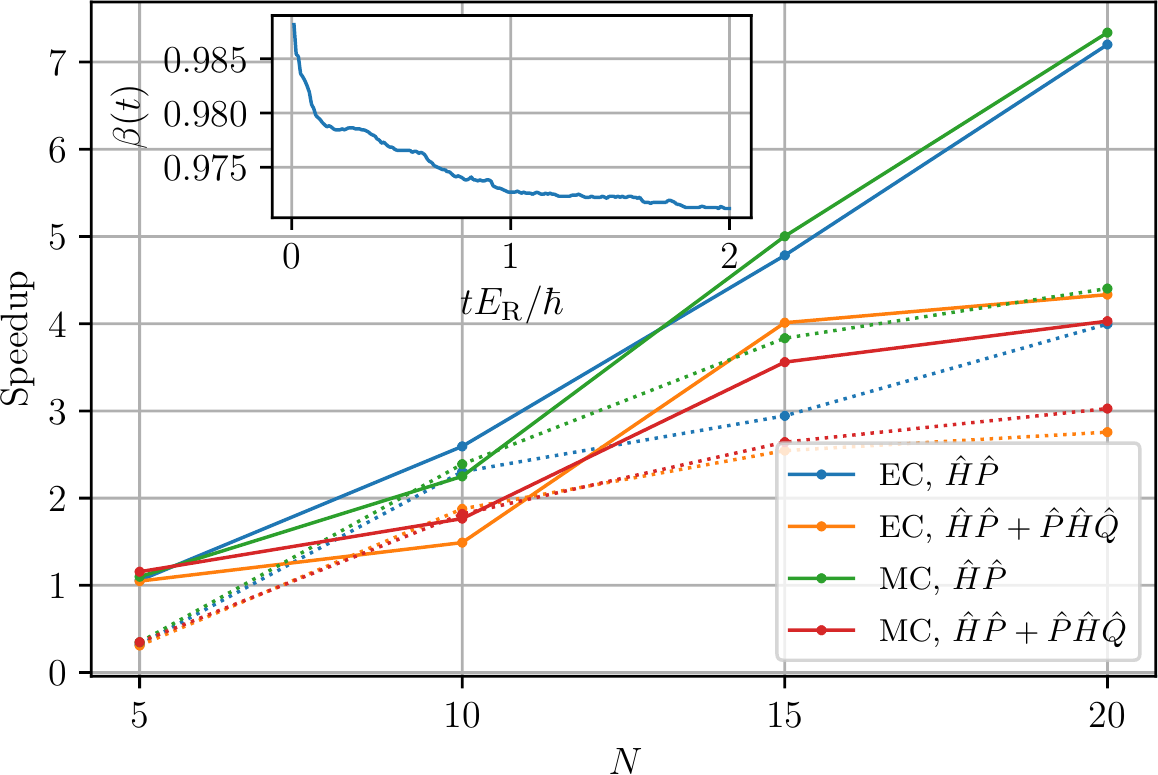}
  \caption{\label{fig:lattice_speedup} Speedup of the pruned compared to the unpruned simulations for the five-well lattice. The solid lines are affiliated with the weak interaction quench to $\tilde{g}=0.4$ while the dotted lines indicate the strong interaction quench to $\tilde{g}=0.8$. The different colors indicate the number state selection criteria (Equation~\eqref{eq:mc} or~\eqref{eq:ec}) and the modified Hamiltonian (Equation~\eqref{eq:hp_phq} or~\eqref{eq:hp}) is used. The inset shows the ratio of the pruned configurations at each time step $\beta(t)$ (see Equation~\eqref{eq:pruning_ratio}) for the case $N=20$ using the energy criterion and the Hamiltonian $\hat{H}\hat{P}$ after a weak interaction quench to $\tilde{g}=0.4$. A pruning threshold $\gamma=10^{-8}$ and pruning time $\tau=10^{-2}\,\hbar/E_{\mathrm{R}}$ was employed.}
\end{figure}

We apply the various pruning methods described in Sections~\ref{sec:pruning} and~\ref{sec:criteria} to this lattice setup and choose a pruning threshold of $\gamma=10^{-8}$ and a pruning time $\tau=10^{-2}\,\hbar/E_{\mathrm{R}}$.
Note that these values of $\gamma$ and $\tau$ have been determined by performing simulations for different sets of parameters $(\gamma,\tau)$.
We find that this combination yields both a significant speed-up compared to the unpruned \gls{mctdhb} while reproducing the unpruned results accurately up to a certain degree.
Figure~\ref{fig:lattice_speedup} shows the reduction of the simulation time in comparison with the regular \gls{mctdhb} for different numbers of particles.
The initial state can be described using only a few configurations such that for small times almost all configurations can be marked as disabled.
Over time this number reduces as can be seen in the inset of Figure~\ref{fig:lattice_speedup}.
This fact can be explained with scattering from the few initially important configurations to the lesser important ones as mediated by the term $\hat{Q}\hat{H}\hat{P}$ in Equations~\eqref{eq:hp_phq} and~\eqref{eq:hp}.
However, the number of inactive configurations remains large throughout the simulation such that a significant speedup compared to the regular \gls{mctdhb} calculation is achieved.
The performance benefit depends on the pruned Hamiltonian that is used as well as the number of particles $N$ (see Figure~\ref{fig:lattice_speedup}).
The evaluation of the pruning criterion introduces computational overhead.
When propagating wave functions containing only few configurations, e.g.\ when studying small particle numbers ($N\approx 5$), only a small speedup can be achieved.
However, for larger systems ($N=20$) a considerably larger speedup by a factor of more than seven can be reached.
The \glspl{eom} based on the Hamiltonian $\hat{H}\hat{P}$ yield higher performance gains than the ones based on $\hat{H}\hat{P}+\hat{P}\hat{H}\hat{Q}$.
This is to be expected as the first variant incorporates less of the original matrix elements while the pruning ratios did not differ substantially.
Furthermore, the stronger quench requires a higher number of configurations and thus leads to a smaller speedup.

\begin{figure}[ht]
  \centering
  \includegraphics[width=0.45\textwidth]{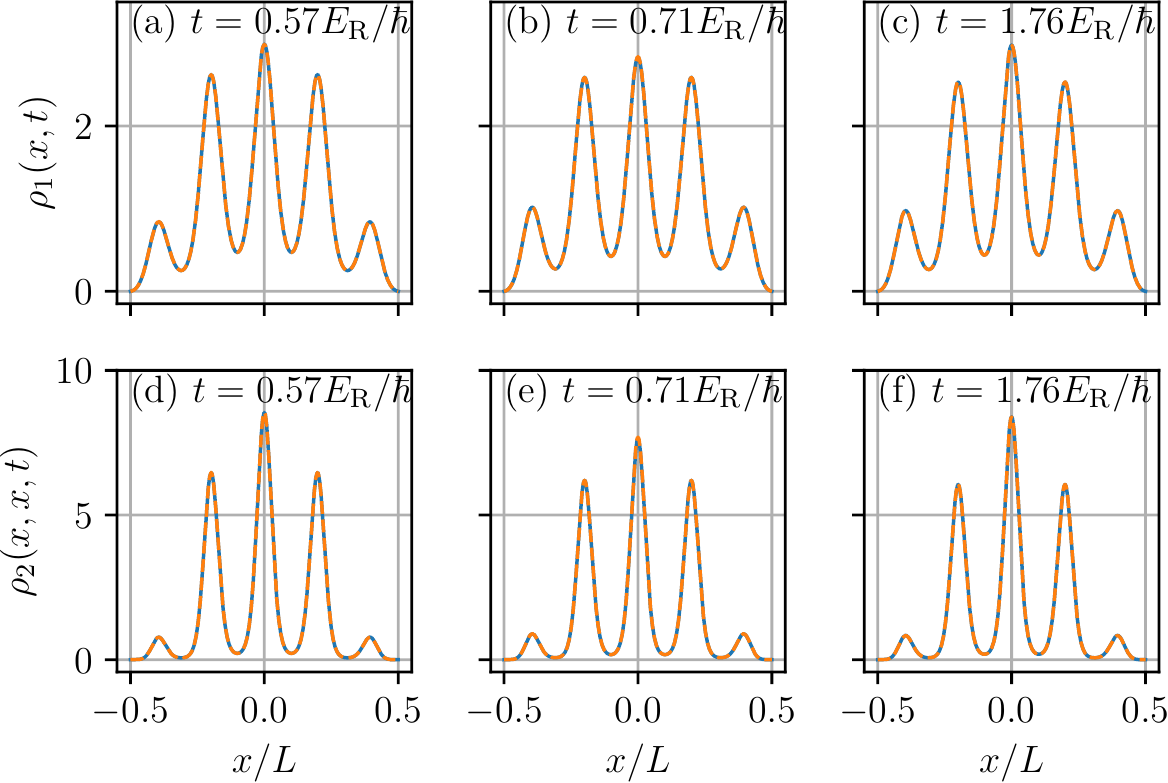}
  \caption{\label{fig:lattice_gpop_snapshots} One-body density $\rho_1(x,t)$ (a)-(c) and diagonal of the reduced two-body density matrix $\rho_2(x,x,t)$ (d)-(f) at selected time instants (see legends) for $N=20$ particles in the five-well lattice following an interaction quench from $\tilde{g}=0.1$ to $\tilde{g}=0.4$. The solid blue lines correspond to the regular \gls{mctdhb} and the dashed orange lines to a pruned calculation using the energy criterion with the Hamiltonian $\hat{H}\hat{P}+\hat{P}\hat{H}\hat{Q}$. However, due to the good agreement of the unpruned and pruned calculation, these lines lie on top of each other. A pruning threshold $\gamma=10^{-8}$ and pruning time $\tau=10^{-2}\,\hbar/E_{\mathrm{R}}$ was employed.}
\end{figure}

Moreover, comparing the \gls{mctdhb} and the pruned one- and two-body density (see Figure~\ref{fig:lattice_gpop_snapshots}) no difference is noticeable throughout the evolution.
Thereby, we can infer that both quantities are reproduced accurately in the pruned simulations.
When applying the \gls{ec} in conjunction with the Hermitian Hamiltonian~\eqref{eq:hp_phq} for $N=20$ bosons and a post-quench interaction of $\tilde{g}=0.4$, the corresponding, maximal absolute deviation is $0.016$ and $0.13$ for the one- and two-body density respectively over the evolution.

\begin{figure}[ht]
  \centering
  \includegraphics[width=0.45\textwidth]{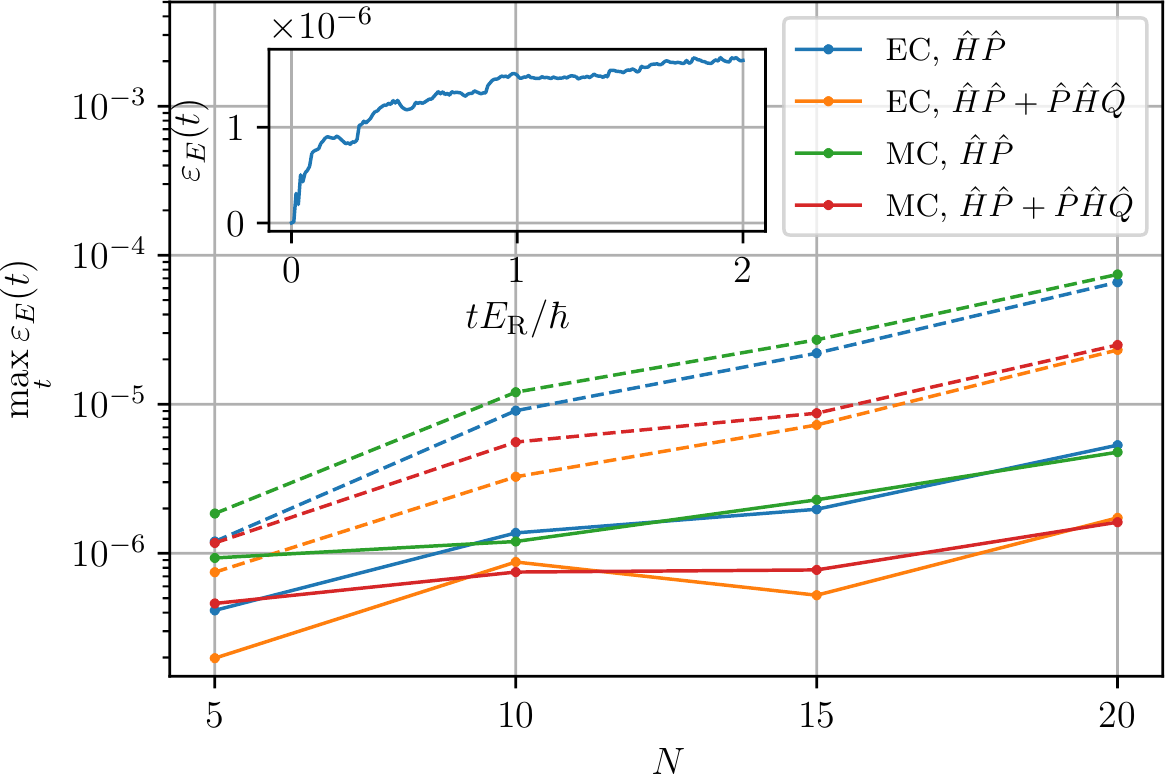}
  \caption{\label{fig:lattice_energy_error} Maximal energetic error (see Equation~\eqref{eq:relative_energy_error}) between the unpruned and pruned simulations for the five-well lattice. The solid lines correspond to a quench to $\tilde{g}=0.4$ and the dashed lines to $\tilde{g}=0.8$. The inset shows the exemplary evolution of the error over time for $\tilde{g}=0.4$ and $N=20$ when using the energy criterion and the Hamiltonian $\hat{H}\hat{P}+\hat{P}\hat{H}\hat{Q}$. A pruning threshold of $\gamma=10^{-8}$ and pruning time of $\tau=10^{-2}\,\hbar/E_{\mathrm{R}}$ was employed.}
\end{figure}

The approximation via the pruning procedure introduces deviations in the energy of the system as well as the natural populations when comparing to the usual \gls{mctdhb}.
The results for the energetic error $\varepsilon_{E}(t)$ (see Equation~\eqref{eq:relative_energy_error}) are illustrated in Figure~\ref{fig:lattice_energy_error}.
As it can be seen, we can reproduce the \gls{mctdhb} energy up to a precision of the order of $10^{-6}$ to $10^{-5}$ for the weaker quench to $\tilde{g}=0.4$ and of the order of $10^{-5}$ to $10^{-4}$ for the stronger quench to $\tilde{g}=0.8$.
The energetic error increases slightly with the number of particles and is higher for the stronger quench.
Among the different pruning approaches only minor differences are perceivable in this quantity.
For instance, the Hermitian operator $\hat{H}\hat{P}+\hat{P}\hat{H}\hat{Q}$ yields slightly smaller errors.
On the other hand, the differences between the energy and the magnitude criterion are negligible.
In conjunction with the significant speedup, we consider these deviations from the \gls{mctdhb} energy to be acceptable.
The inset of Figure~\ref{fig:lattice_energy_error} shows the evolution of $\varepsilon_E(t)$ for $N=20$ particles after a quench to $\tilde{g}=0.4$ for the energy criterion in conjunction with the Hermitian Hamiltonian.
As it can be seen, $\varepsilon_E(t)$ exhibits a fast growth rate initially, while it increases slowly for $tE_{\mathrm{R}}/\hbar\gtrapprox 1$ and tends to saturate.
The long time evolution of $\varepsilon_E(t)$ is discussed for some case examples in Appendix~\ref{sec:long_term}.

\begin{figure}[ht]
  \centering
  \includegraphics[width=0.45\textwidth]{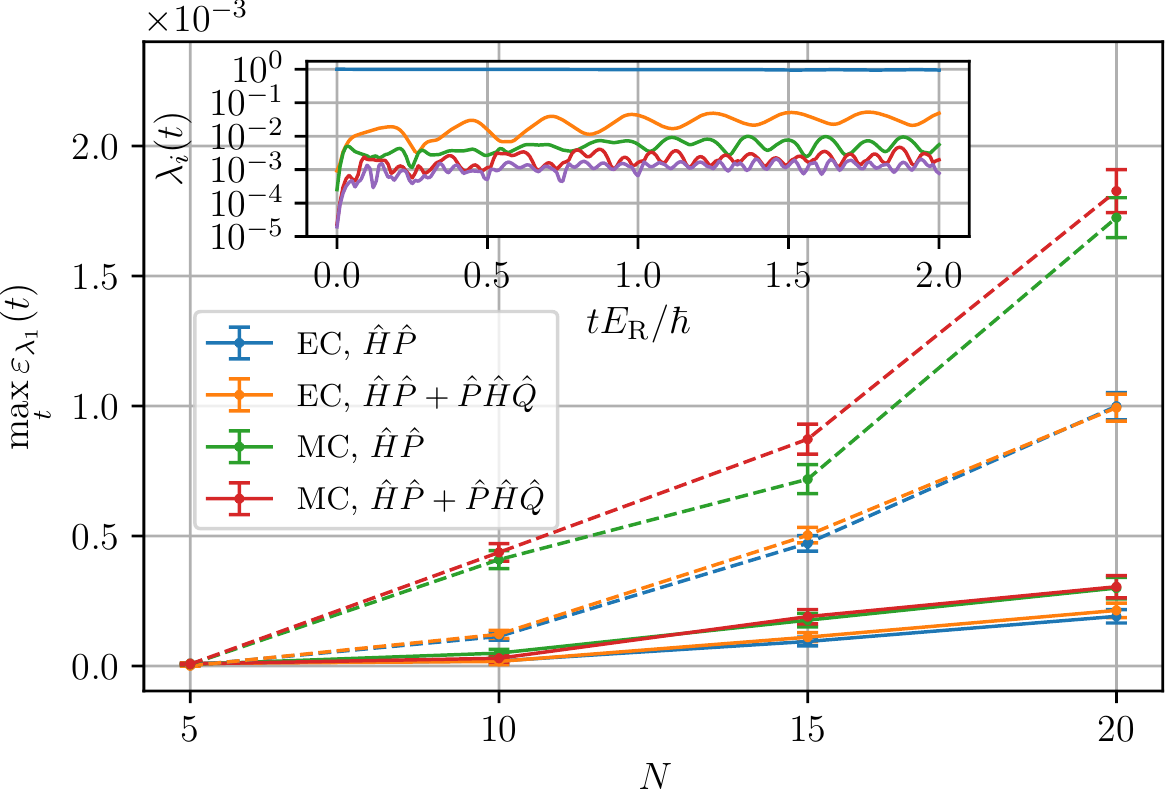}
  \caption{\label{fig:lattice_natpop_error} Maximum error of the first natural population as given by Equation~\eqref{eq:absolute_natpop_error} for various numbers of bosons in a five-well lattice. The solid lines illustrate a post-quench interaction strength of $\tilde{g}=0.4$ while the dashed lines represent $\tilde{g}=0.8$. The error bars indicate the standard deviation of the error. The inset shows the evolution of all $5$ natural populations for the case $N=20$ after a weak quench to $\tilde{g}=0.4$ and using the regular \gls{mctdhb}. A pruning threshold of $\gamma=10^{-8}$ and pruning time of $\tau=10^{-2}\,\hbar/E_{\mathrm{R}}$ was employed.}
\end{figure}

When reviewing the natural populations during the post-quench propagation, we observe that the system cannot be considered a condensed system that could be described in a mean-field manner using only a single orbital as more than one natural orbital is macroscopically occupied.
An example for the dynamical behavior of the natural populations is shown in the inset of Figure~\ref{fig:lattice_natpop_error}.
The maximal depletion over time, i.e.\ the maximal deviation of the first natural population from unity, increases with $N$ and is larger for the stronger quench to $\tilde{g}=0.8$.
For the weak quench to $\tilde{g}=0.4$, we observe a maximal depletion of $\max\limits_t\left(1-\lambda_1(t)\right)=0.07$ for $N=20$ particles.
Similarly, the depletion for a quench to $\tilde{g}=0.8$ exhibits a maximum value of $\max\limits_t\left(1-\lambda_1(t)\right)=0.16$ for $N=20$ particles.

In Figure~\ref{fig:lattice_natpop_error} we also compare the maximal deviation of the natural populations $\max\limits_{t}\varepsilon_{\lambda_i}(t)$ (see Equation~\eqref{eq:absolute_natpop_error}) between the pruned and the unpruned simulations over time.
We exemplarily present the results for the dominant, first orbital.
Over time $\varepsilon_{\lambda_1}(t)$, shows an oscillatory behavior around a central value so that we compute the standard deviation of this quantity to quantify these fluctuations (see error bars in Figure~\ref{fig:lattice_natpop_error}).
For the natural population error, the pruning criterion has a larger impact than the type of \gls{eom} being used.
The energy criterion (blue and orange line) shows slightly better results than the magnitude criterion (green and red line) but stays in the same order of magnitude.
As the energetic error, the error increases with the system size but stays of the order of $10^{-4}$ for a post-quench interaction of $\tilde{g}=0.4$ and of the order of $10^{-3}$ for $\tilde{g}=0.8$.

\begin{figure}[ht]
  \centering
  \includegraphics[width=0.45\textwidth]{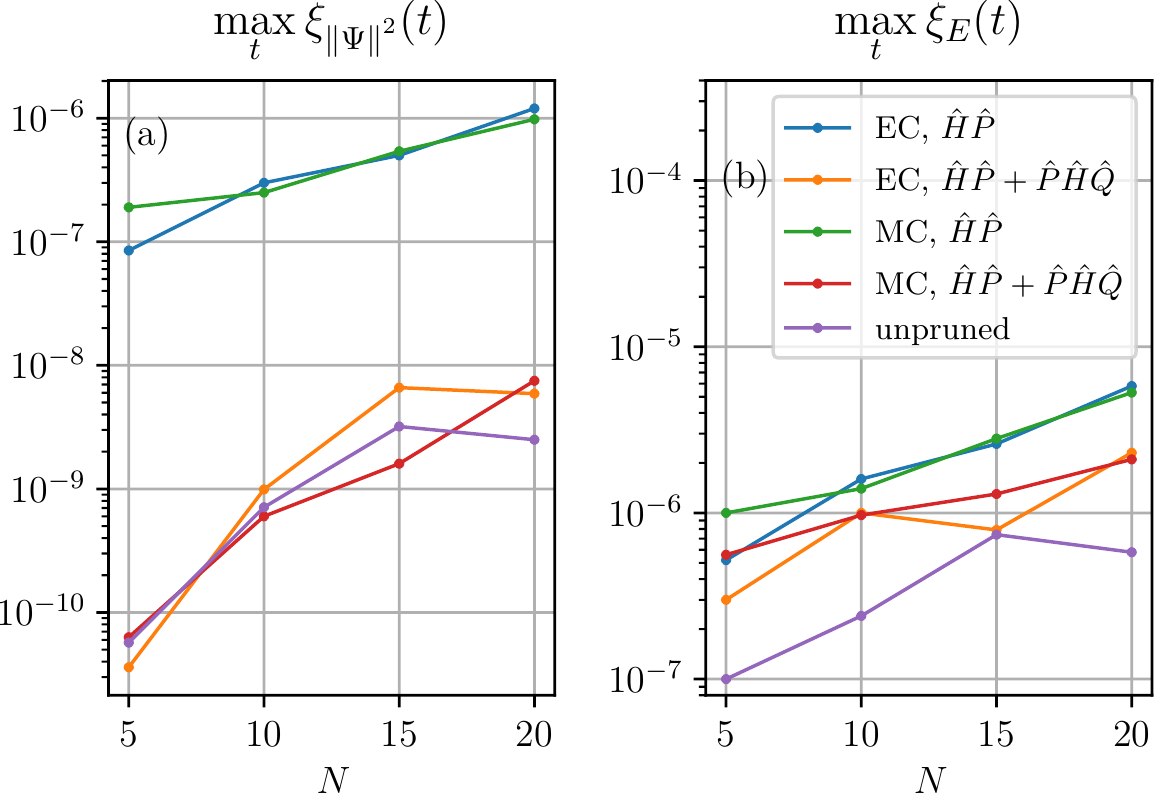}
  \caption{\label{fig:lattice_conservation} Maximal violation of the conservation of (a) the norm (see Equation~\eqref{eq:norm_conservation}) and (b) the energy (see Equation~\eqref{eq:energy_conservation}) during the propagation of the five-well lattice system for an increasing number of particles $N$.}
\end{figure}

The post-quench Hamiltonian is time-independent and therefore the energy should be conserved in addition to the norm of the wave function.
Figure~\ref{fig:lattice_conservation} shows the maximum violation of these constraints over time.
The norm conservation as quantified by $\xi_{{\left\|\Psi\right\|}^2}(t)$ (see Equation~\eqref{eq:norm_conservation}) shows a drastic difference between the two kinds of \glspl{eom}.
The non-Hermitian Hamiltonian leads to a deviation that is a few orders of magnitude larger, while also still remaining sufficiently small.
The violation of the energy conservation $\xi_{E}(t)$ (see Equation~\eqref{eq:energy_conservation}), is slightly higher with the non-Hermitian \glspl{eom}.
Overall however, this error is small and acceptable.
\FloatBarrier{}

\subsection{\label{sec:ho}Nonequilibrium Dynamics in a Double Well}
Our second physical example system is an ensemble of interacting bosons confined in a double-well that is created from a harmonic trap with an additional Gaussian-shaped barrier at the center, also known as a dimple trap in the literature~\cite{akram2016},
\begin{equation}
  V(x)=\frac{1}{2}m\omega² x^2 + V_0\exp\left(-\frac{x^2}{2\sigma^2}\right).
\end{equation}
Here, $\sigma$ is the standard deviation of the Gaussian and $V_0$ is the height of the barrier.
We use the harmonic oscillator length $l_{\mathrm{H}}=\sqrt{\sfrac{m\omega}{\hbar}}$ as the natural length scale of the system with $\omega$ being the angular frequency of the harmonic potential.
The energy units are given by $\hbar\omega$ and the time units by $\sfrac{1}{\omega}$.
In order to induce the nonequilibrium dynamics in this setup we prepare the ground state of the system without a barrier (i.e.\ for $V_0=0$) and then quench the barrier height to $\sfrac{V_0}{\hbar\omega}=4$.
A similar scheme, where the central barrier was continuously ramped up was investigated in~\cite{streltsov2007}.

We study setups with $N=5,10,15,20,25,30$ particles using $m=10,9,7,6,5,5$ orbitals respectively and ensure convergence with respect to $m$ (see Appendix~\ref{sec:convergence}).
The interaction strength is chosen to be $\sfrac{g}{\hbar\omega l_{\mathrm{H}}}=0.1$ and the width of the barrier to be $\sigma=l_{\mathrm{H}}$.
In order to choose the pruning threshold $\gamma$ and the pruning time $\tau$ appropriately, we perform simulations for different values and compared the results to an unpruned \gls{mctdhb} simulation.
We discuss the convergence procedure for different values of $\tau$ and $\gamma$ in detail for the case of $N=15$ particles in Appendix~\ref{sec:pruning_parameters}.
A pruning threshold of $\gamma=10^{-10}$ and a pruning time of $\tau\omega=5\cdot 10^{-2}$ lead to a good agreement with the unpruned~\gls{mctdhb} simulations.

\begin{figure}[ht]
  \centering
  \includegraphics[width=0.45\textwidth]{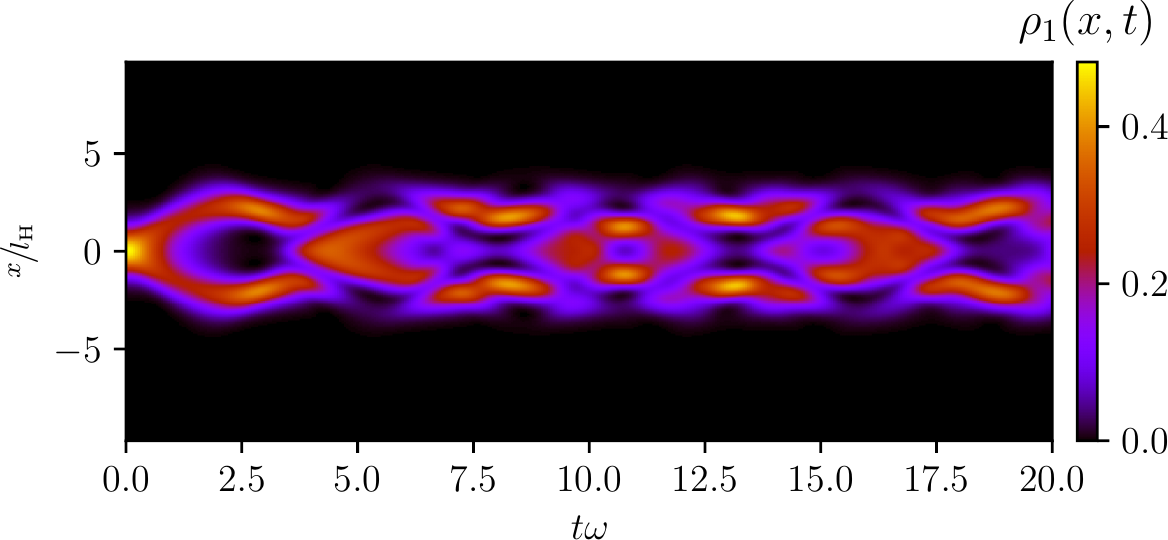}
  \caption{\label{fig:ho_gpop} Time evolution of the one-body density $\rho_1(x,t)$ of $N=15$ bosons in a double-well trap after a quench of the central Gaussian barrier to a finite height obtained with a regular \gls{mctdhb} calculation.}
\end{figure}

The evolution of the single-particle density of the system is showcased in Figure~\ref{fig:ho_gpop}.
By quenching to a finite height of the central barrier, the initial Gaussian distribution of the bosons is split into two branches veering away from each other with opposite momenta.
With the given parameters, the two clouds possess enough energy to overcome the hump after being reflected by the harmonic trap and collide in the trap center $x=0$ at a time $t\omega\approx 4$.
Afterwards, these density branches separate again each one moving in one of the wells of the double-well and subsequently collide at $x=0$ again.
This motion is repeated almost periodically throughout the evolution.
Our main focus is the performance of the pruning approach in this scenario, i.e.\ we do not analyze the overall dynamics in further detail.

\begin{figure}[ht]
  \centering
  \includegraphics[width=0.45\textwidth]{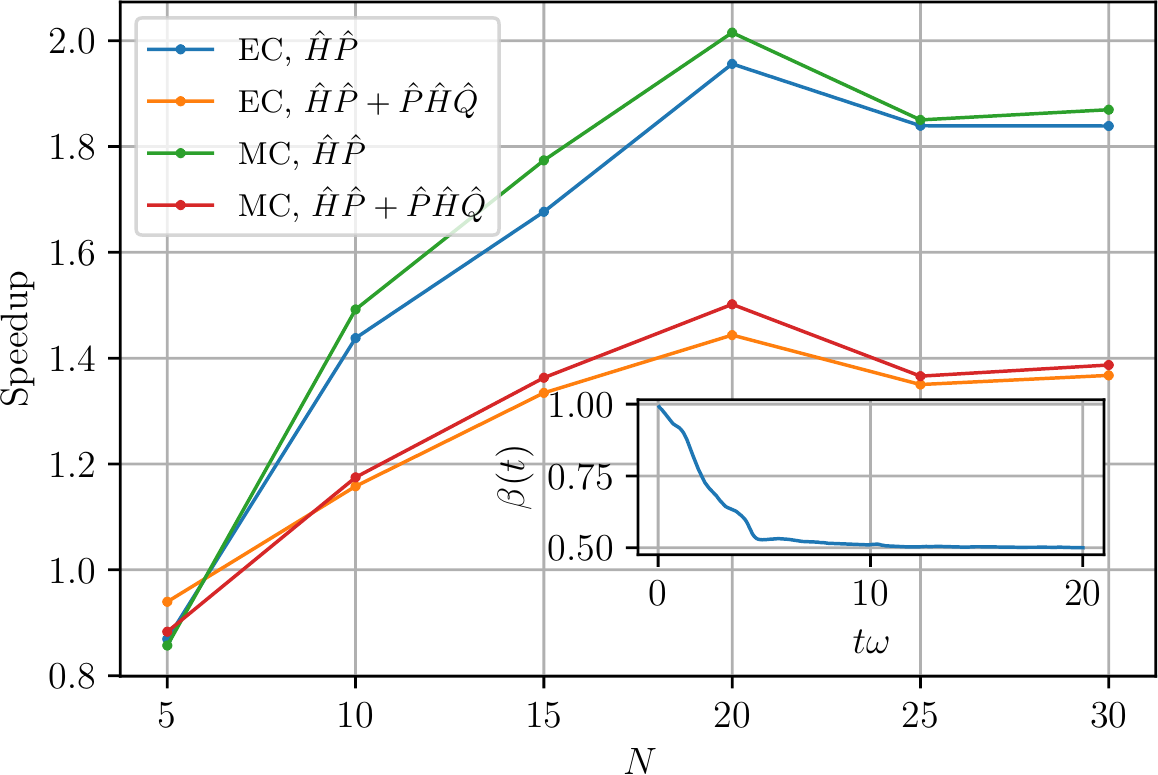}
  \caption{\label{fig:ho_speedup} Speedup of the various pruned simulations compared to regular \gls{mctdhb} in the double-well setup for varying particle numbers. The inset shows the ratio of inactive configurations $\beta(t)$ (see~\eqref{eq:pruning_ratio}) over time for $N=15$ particles using the energy criterion and the Hamiltonian $\hat{H}\hat{P}$. We used a pruning threshold of $\gamma=10^{-10}$ and a pruning time of $\tau\omega=5\cdot 10^{-2}$.}
\end{figure}

Figure~\ref{fig:ho_speedup} shows the speedup of the pruned versus the unpruned simulations.
In comparison to the lattice system, the benefits of the pruning approach are smaller yielding a speedup between $1.4$ and $2$ depending on the system size.
This can be explained by a smaller pruning ratio as it can be seen in the inset.
The ratio of inactive configurations quickly drops from almost $1$ to around $0.5$ where it saturates, suggesting that a higher amount of configurations is required to describe the physical system accurately.

\begin{figure}[ht]
  \centering
  \includegraphics[width=0.45\textwidth]{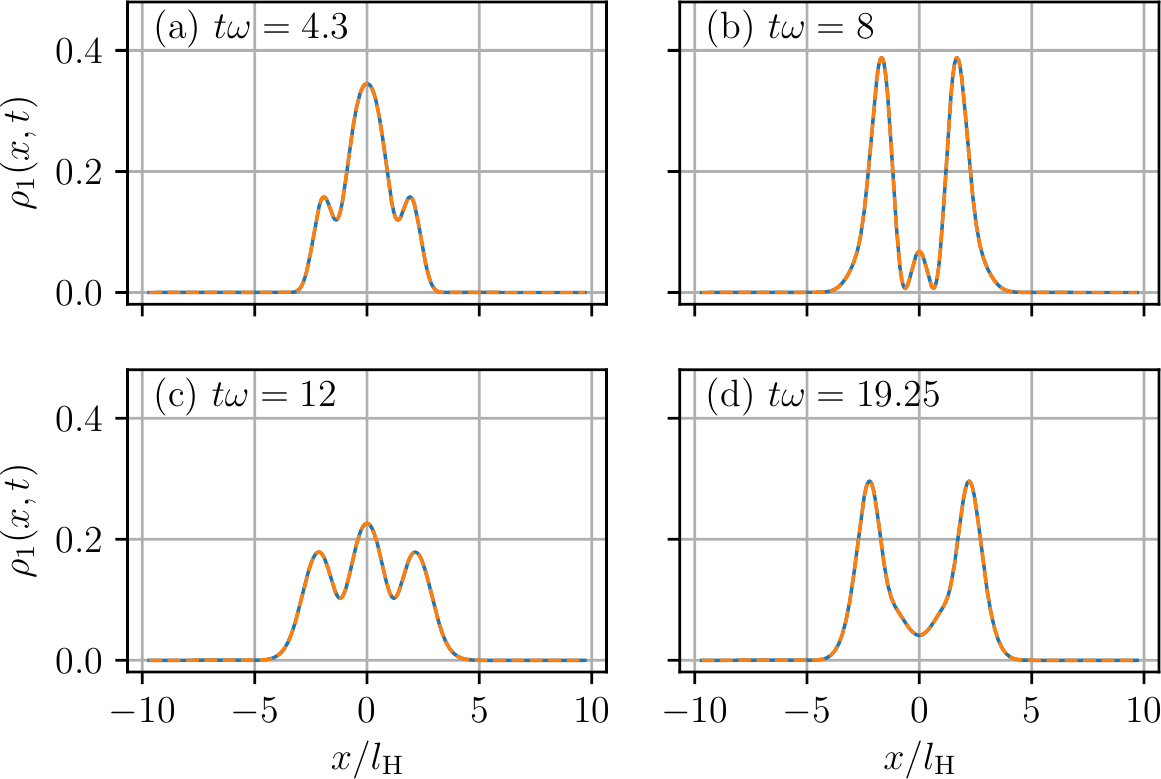}
  \caption{\label{fig:ho_gpop_snapshots} One-body density $\rho_1(x,t)$ for $N=15$ particles in the double-well setup at various time instances (see legends). The blue solid line corresponds to an unpruned simulation and the orange dashed line to a pruned calculation using the energy criterion and the Hamiltonian $\hat{H}\hat{P}+\hat{P}\hat{H}\hat{Q}$. However, due to the good agreement between the unpruned and the pruned calculation these lines lie on top of each other. We used a pruning threshold of $\gamma=10^{-10}$ and a pruning time of $\tau\omega=5\cdot 10^{-2}$.}
\end{figure}

\begin{figure}[ht]
  \centering
  \includegraphics[width=0.45\textwidth]{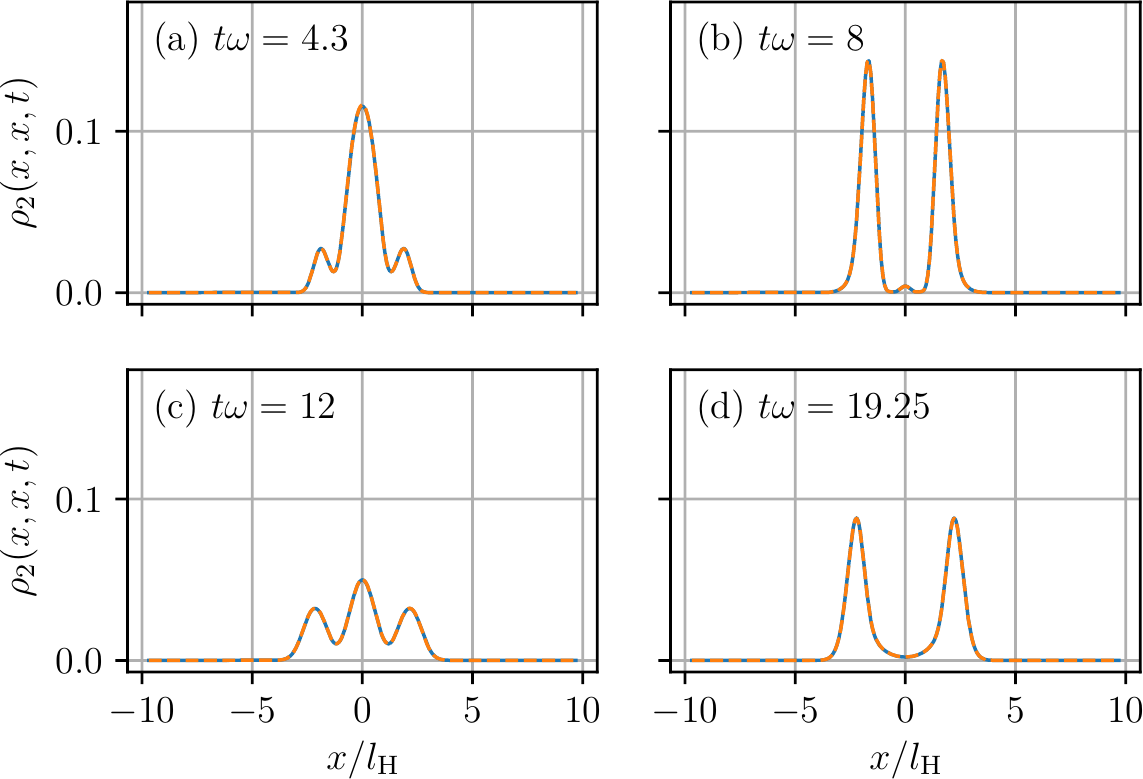}
  \caption{\label{fig:ho_dmat2_diagonal} Comparison of the diagonal $\rho_2(x,x^\prime,t)$ of the two-body density matrix at different times $t$ (see legends) for $N=15$ particles in a double-well following the quench of the central Gaussian barrier to a finite height. The solid blue line corresponds to a regular \gls{mctdhb} simulation and the orange dashed line to a pruned calculation using the \gls{ec} and the $\hat{H}\hat{P}+\hat{P}\hat{H}\hat{Q}$ Hamiltonian. However, due to the good agreement between the unpruned and the pruned calculation these lines lie on top of each other. We used a pruning threshold of $\gamma=10^{-10}$ and a pruning time of $\tau\omega=5\cdot 10^{-2}$.}
\end{figure}

Even though the pruning approach does not speed up the simulation as much as in the case of an optical lattice, the evolution of the system is still reproduced accurately.
We show the good agreement of the one-body (see Figure~\ref{fig:ho_gpop_snapshots}) and the two-body (see Figure~\ref{fig:ho_dmat2_diagonal}) densities between a regular \gls{mctdhb} and a pruned simulation using the \gls{ec} in conjunction with the Hermitian Hamiltonian~\eqref{eq:hp_phq}.
The corresponding maximal, absolute deviation over the evolution time and the whole grid is $0.0013$ for the one-body and $0.0012$ for the two-body density.

\begin{figure}[ht]
  \centering
  \includegraphics[width=0.45\textwidth]{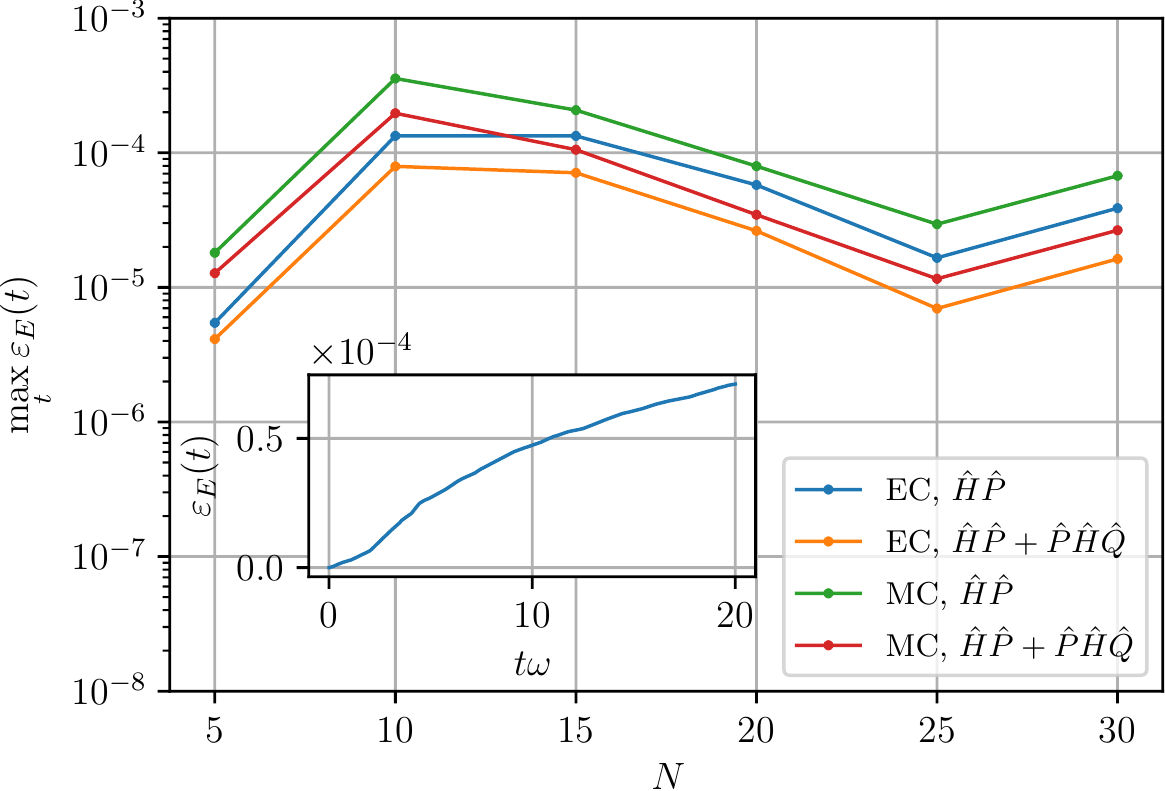}
  \caption{\label{fig:ho_energy_error} Maximal relative energetic error (see Eq.~\eqref{eq:relative_energy_error}) between unpruned and pruned simulations for the double-well setup for varying particle number $N$. The inset shows the evolution of the error over time during a pruned simulation for $N=15$ particles using the energy criterion and the $\hat{H}\hat{P}+\hat{P}\hat{H}\hat{Q}$ Hamiltonian. We used a pruning threshold of $\gamma=10^{-10}$ and a pruning time of $\tau\omega=5\cdot 10^{-2}$.}
\end{figure}

The energetic error $\varepsilon_{E}$ (see Equation~\eqref{eq:relative_energy_error}) between the pruned and the unpruned \gls{mctdhb} calculation is depicted in Figure~\ref{fig:ho_energy_error}.
In contrast to the lattice system, the difference between the two types of \glspl{eom}~\eqref{eq:eom_hp_phq} and~\eqref{eq:eom_hp} is only minor and the error does not increase with the number of particles.
Using the pruning approach, the unpruned \gls{mctdhb} energy is reproduced up to a relative deviation of the order of $10^{-4}$.
The inset of Figure~\ref{fig:ho_energy_error} shows that the energetic error grows in a similar fashion as in the lattice system over the simulated time and that it does saturate within the given time range.
In Appendix~\ref{sec:long_term} we show exemplarily the long-time evolution of the energetic error for fixed particle number and observe that it grows slowly at longer propagation times.

\begin{figure}[ht]
  \centering
  \includegraphics[width=0.45\textwidth]{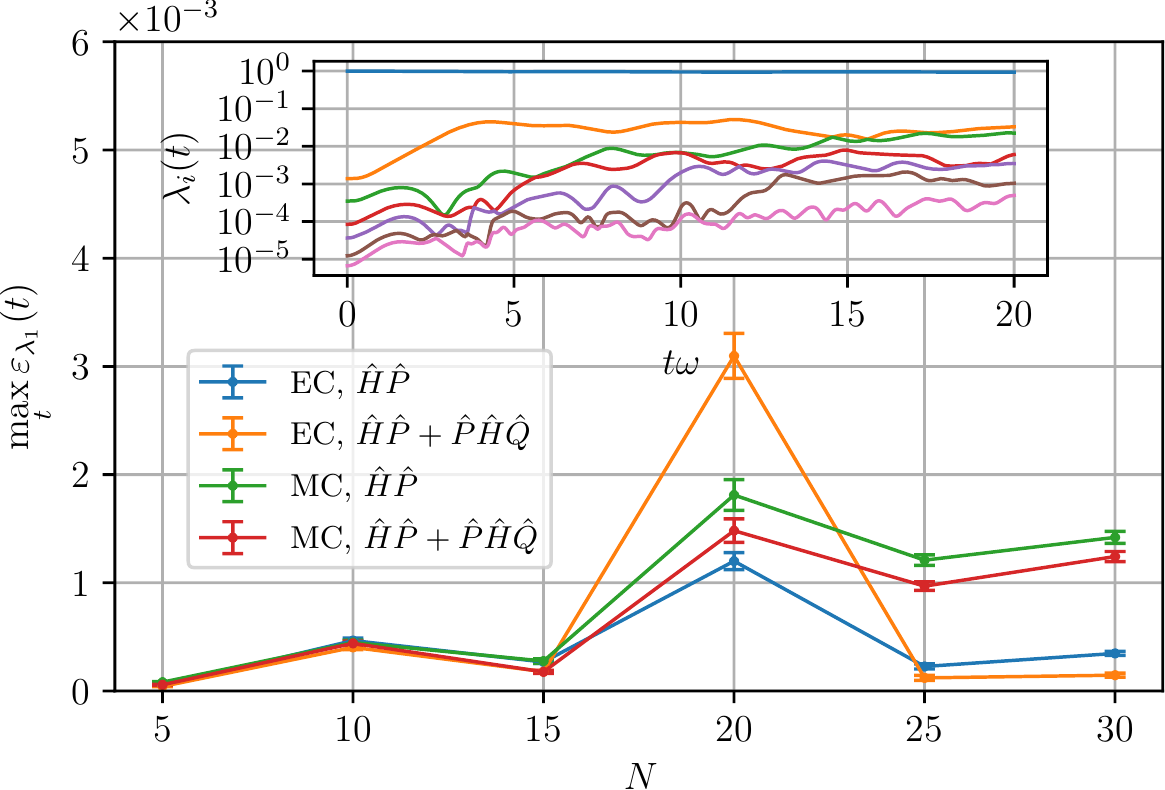}
  \caption{\label{fig:ho_natpop_error} Absolute error of the first natural population (see Eq.~\eqref{eq:absolute_natpop_error}) between the pruned and the unpruned simulations in the double-well setup. The error bars indicate the standard deviation of this error quantity. The inset shows the evolution of all $m=7$ orbitals for the unpruned \gls{mctdhb} calculation using $N=15$ particles. We used a pruning threshold of $\gamma=10^{-10}$ and a pruning time of $\tau\omega=5\cdot 10^{-2}$.}
\end{figure}

Again, we also investigate the impact of the pruning on the natural orbitals.
The maximal absolute error of the first, dominant natural population $\max\limits_t \varepsilon_{\lambda_1}(t)$ (see Equation~\eqref{eq:absolute_natpop_error}) is shown in Figure~\ref{fig:ho_natpop_error}.
We observe that it is at most of the order of $10^{-3}$ verifying that the first natural population is reproduced accurately.
As in the lattice system, the depletion of the system increases with the number of particles.
For $N=30$ particles we achieve a maximal depletion of $\max\limits_t\left(1-\lambda_1(t)\right)=0.125$ so that the given parameters lead to beyond-mean-field dynamics.
The evolution of the natural populations with time is visualized in the inset of Figure~\ref{fig:ho_natpop_error}.

\begin{figure}[ht]
  \centering
  \includegraphics[width=0.45\textwidth]{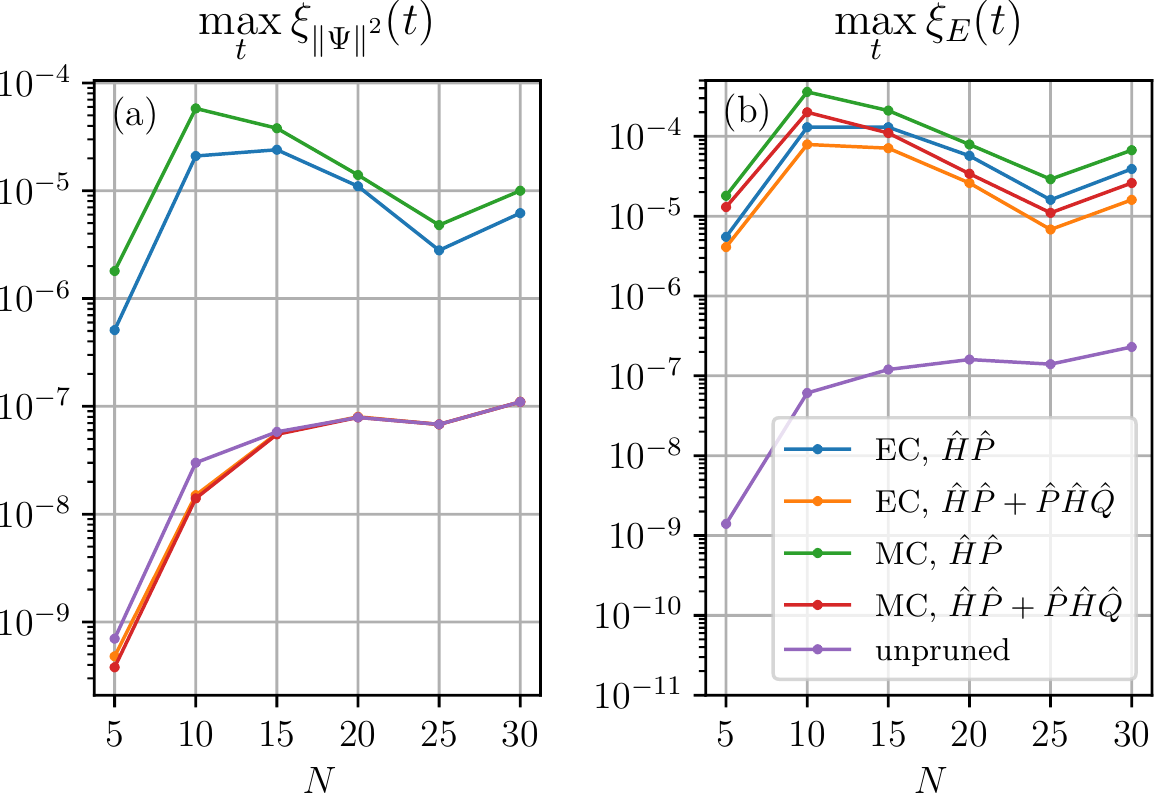}
  \caption{\label{fig:ho_conservation} Maximal violation of the conservation of (a) the norm (see Eq.~\eqref{eq:norm_conservation}) and (b) the energy (see Eq.~\eqref{eq:energy_conservation}) during the propagation after quenching the central Gaussian barrier in the double-well to a finite height. We use a pruning threshold of $\gamma=10^{-10}$ and a pruning time of $\tau\omega=5\cdot 10^{-2}$.}
\end{figure}

In Figure~\ref{fig:ho_conservation} (a) we show the violation of the norm conservation $\xi_{{\left\|\Psi\right\|}^2}(t)$ (see Equation~\eqref{eq:norm_conservation}) for our double-well setup.
Similarly to the lattice system, we see a discrepancy between the two types of \glspl{eom} with the Hermitian operator $\hat{H}\hat{P}+\hat{P}\hat{H}\hat{Q}$ yielding comparable results to the unpruned calculations and the non-Hermitian operator $\hat{H}\hat{P}$ producing errors that are a few orders of magnitude higher.
After the quench, the Hamiltonian is time-independent such that the total energy should be conserved.
Again, we observe a deviation of this law $\xi_{E}(t)$ (see Equation~\eqref{eq:energy_conservation}) that is a couple of orders of magnitude higher than in the unpruned simulation while remaining very small overall.
\FloatBarrier{}

\section{\label{sec:conclusions}Conclusions and Outlook}
Studying the nonequilibrium quantum dynamics of large many-body systems poses a great challenge for numerical methods due to the excessively growing number of configurations.
We have presented an intuitive, novel approach to address this issue in the framework of \glsentryfull{mctdhb}.
Our scheme dynamically classifies number states according to their importance for the physical system under consideration employing pruning criteria that can be controlled using tunable accuracy parameters.
We have derived two such criteria based on understandable quantities that can be computed efficiently.
Our approach is dynamical and can adapt the number state selection during the evolution of the system to ensure an accurate description.
The resulting, time-dependent selection of important configurations can be exploited by modifying the \gls{mctdhb} \glspl{eom}.
Our algorithm cannot overcome the exponential growth of the Hilbert space but can greatly reduce the numerical effort by purposefully neglecting terms of the Hamiltonian.

We have benchmarked our scheme using the quench dynamics of two typical systems from the field of ultracold atoms, namely an optical lattice and a double-well.
The dynamical pruning approach is able to accurately reproduce the results of the unpruned \gls{mctdhb}, while often reducing the computation time significantly.
The speedup was particularly large for the lattice system since a large number of coefficients are of minor importance.
The computational gain is much smaller in the double-well setup, suggesting a strong dependence on the physical system under investigation.
In this sense, we are hesitant to universally recommend one of the pruning criteria or one of the modified \glspl{eom} since all choices lead to an accurate description of the unpruned \gls{mctdhb} results.
Therefore, it is worthwhile to study all combinations as this situation might change when investigating new, different physical systems.
In particular when choosing one of the modified Hamiltonians, a tradeoff exists between the accuracy and speedup.
The Hamiltonian $\hat{H}\hat{P}$ (see Eq.~\eqref{eq:hp}) takes fewer of the original matrix elements into account which yields a larger speedup while also introducing additional inaccuracies to the simulations.
In general, the non-hermiticity could be problematic when studying different physical systems and should be checked carefully.
On the other hand, the Hamiltonian $\hat{H}\hat{P}+\hat{P}\hat{H}\hat{Q}$ leads to a better agreement with the unpruned \gls{mctdhb} but offers a smaller speedup as more matrix elements are taken into account.
In terms of the pruning criterion, we observe comparable errors introduced by the pruning approach and no difference in the achievable speed-up.
In some observables, the energy criterion leads to slightly smaller inaccuracies which are not large enough to lead to a general recommendation especially as this situation might be different when studying other setups.

Based on these results, we can conclude that our scheme captures the important aspects of the physical system correctly while reducing the numerical effort, making it an attractive candidate for future investigations.
A promising prospect in doing so is the realization of extrapolation studies.
By studying a physical system both with unpruned and pruned \gls{mctdhb} up to a certain, feasible size one can ensure the agreement of both approaches and that the parameters $\gamma$ and $\tau$ are chosen appropriately.
Afterwards, larger system sizes, that are not achievable using unpruned \gls{mctdhb}, could be investigated using the pruning approach while extrapolating the quantities that have been used to compare to the regular \gls{mctdhb} for the smaller sizes.
Furthermore, the method we presented in this work may be further refined by employing alternative pruning criteria or by modifying the \glspl{eom} in a different manner.
Another promising direction for further studies is the application of the dynamical pruning scheme to other methods from the family of \gls{mctdh} such as the \gls{mctdhf}~\cite{zanghellini2003,caillat2005} for fermionic systems.
Due to the strong interest that developed in the investigation of binary mixtures using \gls{ml-mctdhx}~\cite{kronke2013,cao2013,cao2017} in recent years, the implementation of a dynamical pruning scheme for this method could be very helpful in order to reduce the numerical effort of these time-consuming simulations.
One possible way is to apply the pruning approach presented in this article on a per-species basis.

\begin{acknowledgments}
  The authors thank H. R. Larsson for fruitful discussions.
  F.K.\ and P.S.\ gratefully acknowledge funding by the excellence cluster ``The Hamburg Centre for Ultrafast Imaging -- Structure, Dynamics and Control of Matter at the Atomic Scale'' of the Deutsche Forschungsgemeinschaft.
  S.M.\ and P.S.\ gratefully acknowledge funding by the Deutsche Forschungsgemeinschaft (DFG) in the framework of SFB 925 (``Light induced dynamics and control of correlated quantum systems'').
  K.K.\ acknowledges a scholarship of the Studienstiftung des deutschen Volkes.
\end{acknowledgments}

\appendix

\section{\label{sec:convergence}Convergence of the MCTDHB Calculations}
The \glspl{spf} used by \gls{mctdhb} are variationally optimal, however the number of these orbitals has to be sufficiently large to ensure the numerical exactness of the method.
In order to ensure the convergence with respect to the number of orbitals, we performed calculations with varying number of orbitals.
By comparing the results for different basis sizes, we ensure that the employed observables such as the particle densities do not change up to a certain degree when using more orbitals than the numbers we presented in the main text.
Additionally, the natural populations are important when discussing the convergence of \gls{mctdhb}.
In a converged calculation, the natural populations should show a rapidly decreasing hierarchy and orbitals that are neglected should only be weakly occupied.

We use $m=5$ orbitals for the investigations of the lattice system in Section~\ref{sec:lattice}.
The least occupied orbital shows a maximal natural population of $\max\limits_t\lambda_5(t)=\mathcal{O}(10^{-3})$ for all particle numbers and post-quench interaction strengths throughout the time evolution.
Any orbitals added to the simulation are only weakly occupied.
We observe a clear drop in the natural populations as already the next orbital shows an occupation of $\max\limits_t\lambda_6(t)=\mathcal{O}(10^{-5})$ and further natural populations are even smaller.
In general, the occupation of the last orbital increases with the number of particles and is larger for the stronger quench to $\tilde{g}=0.8$ but only slightly.
Overall, we consider $m=5$ orbitals to be sufficient due to the clear drop in natural populations and the observation that the evolution of the one- and two-body densities does not change qualitatively.
Furthermore, the energy of the final state of the propagation is converged to a precision of at least $\mathcal{O}(10^{-5})$.

For the setup with the double-well presented in Section~\ref{sec:ho}, we used different numbers of orbitals depending on the number of particles.
We ensure that the least occupied orbital that is taken into account is occupied with a natural population of $\max\limits_t\lambda_m(t)=\mathcal{O}(10^{-4})$.
Further orbitals added do not change the behavior of the system qualitatively and the corresponding natural populations decay rapidly.
Additionally, the energy of the final state is converged to at least $\mathcal{O}(10^{-4})$ such that we consider the used number of orbitals to be sufficient.

\section{\label{sec:long_term}Long-Time Evolution of the Energy Error}
\begin{figure}[ht]
  \centering
  \includegraphics[width=0.45\textwidth]{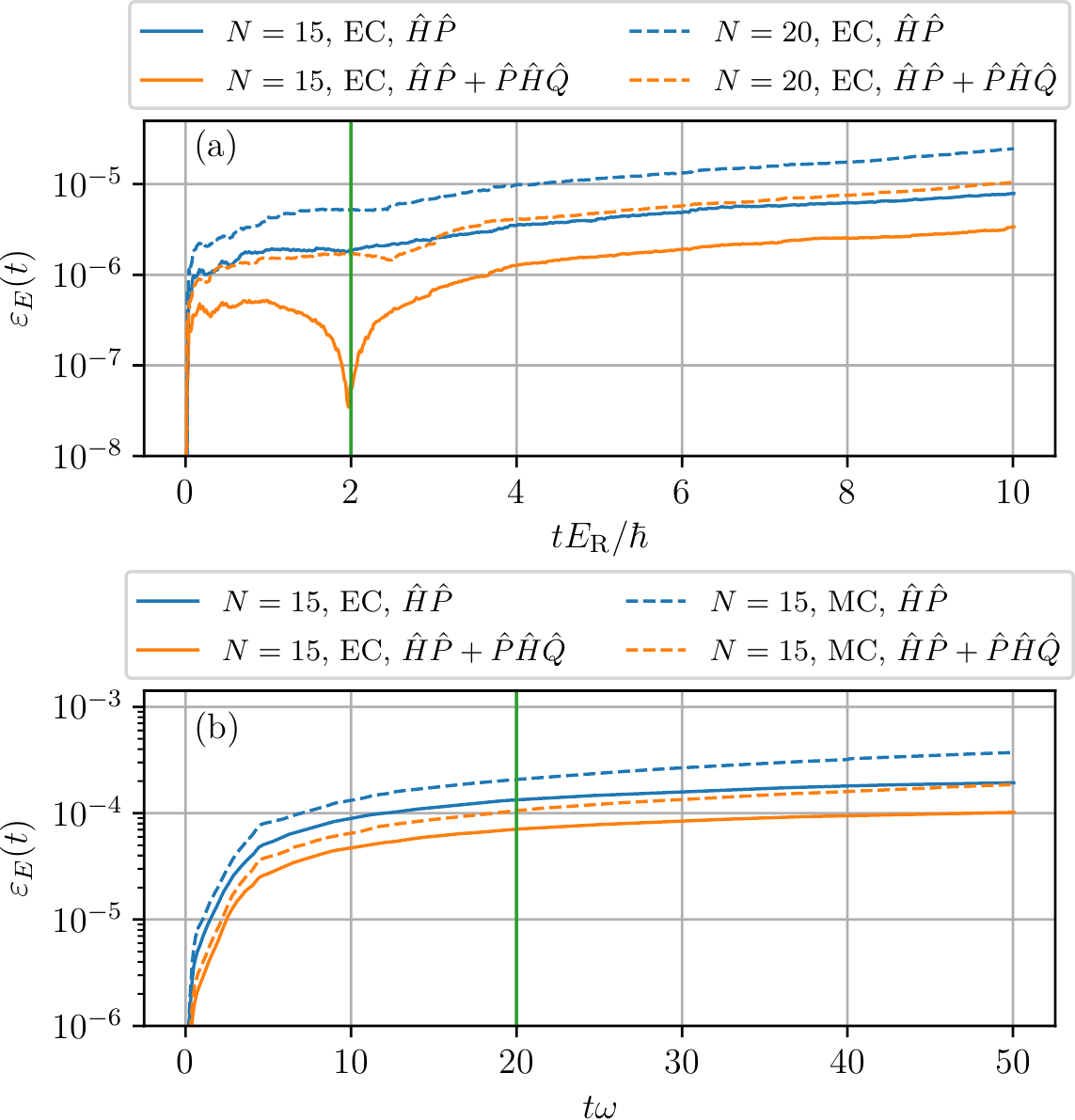}
  \caption{\label{fig:long_term_energy_error}{%
    Time evolution of the relative energetic error $\varepsilon_E(t)$ (see Eq.~\eqref{eq:relative_energy_error}).
    (a) Shows $\varepsilon_E(t)$ for $N=15$ and $N=20$ particles in a five-well lattice following an interaction quench from $\tilde{g}=0.1$ to $\tilde{g}=0.4$.
    A longer propagation time of $t_{\mathrm{f}}=10\hbar/E_{\mathrm{R}}$ compared to Sec.~\ref{sec:lattice} is shown (where $t_{\mathrm{f}}=2\hbar/E_{\mathrm{R}}=2$ as indicated by the green line).
    We use the energy criterion, the Hamiltonian $\hat{H}\hat{P}+\hat{P}\hat{H}\hat{Q}$, a pruning threshold of $\gamma=10^{-8}$ and a pruning time of $\tau=10^{-2}\hbar/E_{\mathrm{R}}$.
    (b) Illustrates $\varepsilon_E(t)$ for $N=15$ bosons in the double-well setup after quenching the central Gaussian barrier to a finite height of $V_0=4\hbar\omega$.
    Compared to the final time of $t_{\mathrm{f}}\omega=20$ in Sec.~\ref{sec:ho} (as indicated by the green line), a longer propagation time $t_{\mathrm{f}}\omega=50$ was used.
    We employ both the energy and the magnitude criterion as well as both modified Hamiltonians (see legend) using a pruning threshold of $\gamma=10^{-10}$ and a pruning time of $\tau\omega=5\cdot 10^{-2}$.
  }}
\end{figure}

In Section~\ref{sec:lattice} we employed a final time of $t_{\mathrm{f}}=2\hbar/E_{\mathrm{R}}$ when studying the lattice setup.
Here, we show the long-time behavior of our pruning approach, i.e.\ we propagate to a final time of $t_{\mathrm{f}}=10\hbar/E_{\mathrm{R}}$, for $N=15$ and $N=20$ particles in a five-well setup following an interaction quench from $\tilde{g}=0.1$ to $\tilde{g}=0.4$ and using the energy criterion and both modified Hamiltonians while propagating.
Figure~\ref{fig:long_term_energy_error} (a) presents the evolution of the corresponding relative energy error $\varepsilon_E(t)$ (see Eq.~\eqref{eq:relative_energy_error}).
The initial wave function at $t=0$ is identical for the pruned and the unpruned simulations, such that initially $\varepsilon_E(0)=0$.
In a short initial time range $tE_{\mathrm{R}}/\hbar$ not much greater than $0$, $\varepsilon_E(0)$ quickly jumps to a small finite value of the order of $10^{-6}$ or $10^{-5}$.
For larger times $t\gtrapprox 4\hbar/E_{\mathrm{R}}$ however, $\varepsilon(t)$ grows only slowly with time, almost saturating, i.e.\ remaining at the same order of magnitude.
In this spirit, our pruning approach is also applicable for the investigation of longer propagation times.

In Section~\ref{sec:ho} we used a final time of $t_{\mathrm{f}}\omega=20$ when investigating the double-well setup.
Here, we showcase the long-time behavior of our pruning approach for $N=15$ particles upon quenching the central Gaussian barrier to a finite height $V_0=4\hbar\omega$ using both pruning criteria and both modified Hamiltonians by propagating to a final time of
$t_{\mathrm{f}}\omega=50$.
In Figure~\ref{fig:long_term_energy_error} (b) we show the evolution of the relative energy error $\varepsilon_E(t)$ (see Eq.~\eqref{eq:relative_energy_error}).
At $t\omega=0$ the pruned and the unpruned simulations coincide, namely $\varepsilon_E(0)=0$.
At small initial times $t\omega\lessapprox 10$ $\varepsilon_E(t)$ grows significantly due to the inaccuracies introduced by the pruning approach but still acquires small values of the order of $10^{-4}$ or $10^{-5}$.
We observe that for large times $t\omega\gtrapprox 10$, $\varepsilon_E(t)$ grows in a slow manner and remains at the same order of magnitude.
Consequently, the pruning approach is also suitable to study longer propagation times of this system.

\section{\label{sec:pruning_parameters}Analysis of the Pruning Parameters}
\begin{figure}[ht]
  \centering
  \includegraphics[width=0.45\textwidth]{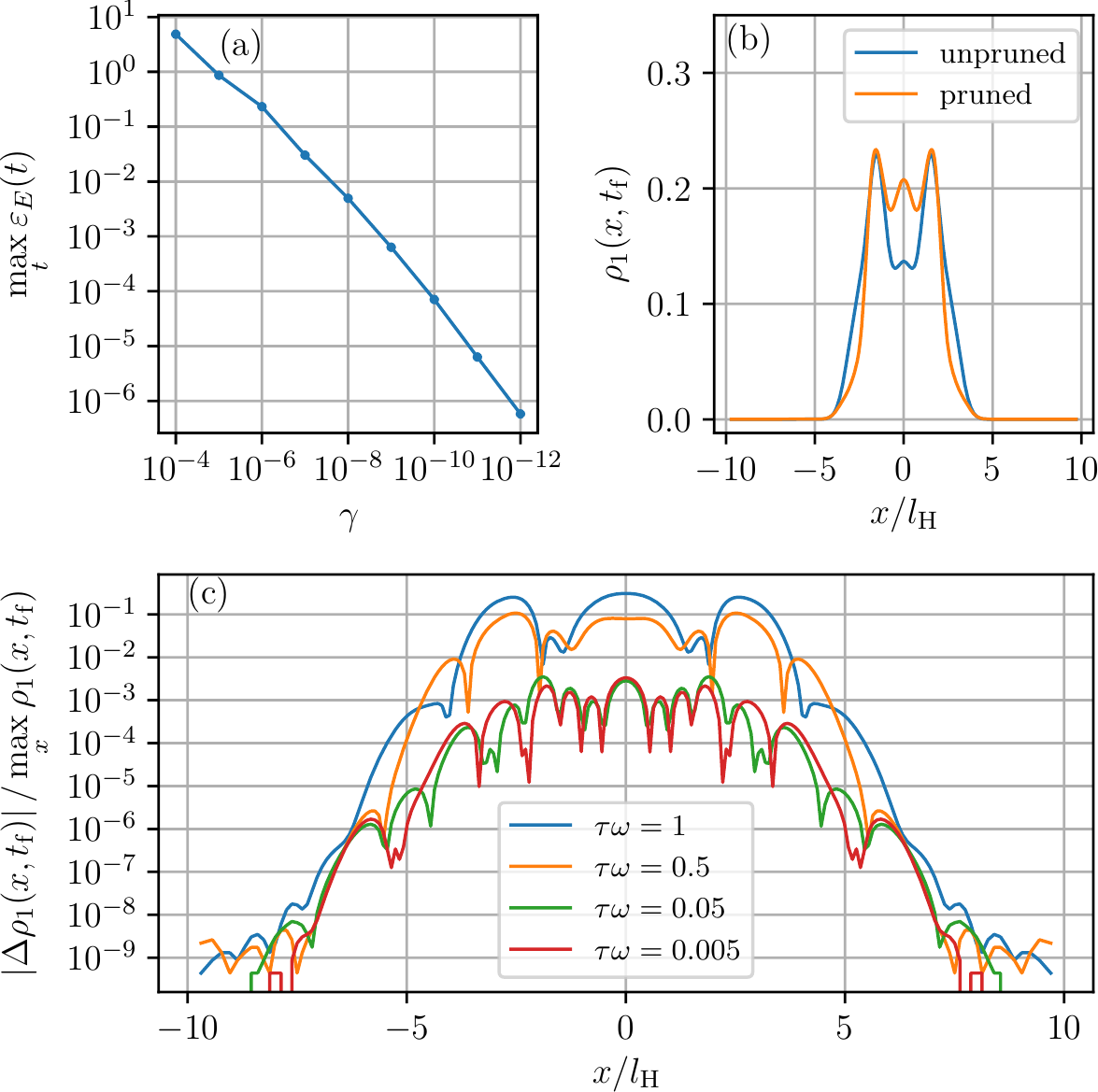}
  \caption{\label{fig:ho_parameters}
    (a) Maximal relative energetic error (see Eq.~\eqref{eq:relative_energy_error}) for varying pruning threshold $\gamma$.
    The pruning time is fixed at $\tau\omega=5\cdot 10^{-2}$.
    (b) One-body density $\rho_1(x,t_{\mathrm{f}})$ after a propagation to a final time of $t_{\mathrm{f}}\omega=20$ using an unpruned (blue line) and a pruned (orange line) \gls{mctdhb} calculation.
    The pruning threshold was chosen as $\gamma=10^{-10}$ and the pruning time as $\tau\omega=1$.
    (c) Error in one-body density at the final time $t_{\mathrm{f}}\omega=20$ when comparing pruned calculations to an unpruned \gls{mctdhb} simulation using varying pruning times $\tau$.
    The quantity shown here is the absolute difference $\left|\Delta\rho_1(x,t_{\mathrm{f}})\right|=\left|\rho_1^\prime(x,t_{\mathrm{f}})-\rho_1(x,t_{\mathrm{f}})\right|$ between the density $\rho_1^\prime(x,t_{\mathrm{f}})$ of the pruned and the density $\rho_1(x,t_{\mathrm{f}})$ of the unpruned simulation normalized to the maximum one-body density.
    For all three figures, the system consists of $N=15$ bosons in a double-well setup (see Sec.~\ref{sec:ho}) following a quench of the central Gaussian barrier to a finite height of $V_0=4\hbar\omega$ for different values of $\tau$ and $\gamma$ using the energy criterion and the Hermitian Hamiltonian $\hat{H}\hat{P}+\hat{P}\hat{H}\hat{Q}$.
  }
\end{figure}

In Sections~\ref{sec:lattice} and~\ref{sec:ho}, we employed fixed values of $\gamma$ and $\tau$ that have been determined by comparing pruned simulations with full \gls{mctdhb} results.
Here, we discuss the impact of these parameters on the accuracy of our pruning approach based on the example of $N=15$ particles in the double-well setup from Section~\ref{sec:ho}.
Figure~\ref{fig:ho_parameters} (a) illustrates the maximal relative energetic error $\max\limits_{t}\varepsilon_{E}(t)$ (see Eq.~\eqref{eq:relative_energy_error}) for a varying pruning threshold $\gamma$ while keeping the pruning time fixed at a value of $\tau=5\cdot10^{-2}$.
We expect that the pruned simulations converge towards the unpruned \gls{mctdhb} results when decreasing $\gamma$.
Indeed, according to our numerical results, the maximal energetic error decreases roughly polynomially with $\gamma$, i.e.\ $\max\limits_t\varepsilon_{E}(t)=b\gamma^k$.
The parameters $k=0.863\pm 0.026$ and $b=\left(2.5\pm 1.3\right)\cdot 10^{4}$ have been determined using a least-squares fit.

Furthermore, we investigate the impact of the pruning time $\tau$ in a similar manner by performing pruned simulations for different values of $\tau$ while keeping the pruning threshold fixed at $\gamma=10^{-10}$.
As shown in Fig.~\ref{fig:ho_parameters} (b), a large value of $\tau$ leads to incorrect results, i.e.\ a discrepancy between the pruning aproach and unpruned~\gls{mctdhb}.
In this example, we show how a value of $\tau\omega=1$ leads to a different final one-body density compared to the unpruned~\gls{mctdhb} which manifests itself in a different shape of the outer flanks and in particular the central peak of the density.
In Fig.~\ref{fig:ho_parameters} (c), we show the error in the final one-body density when using various values for $\tau$ and a fixed pruning threshold of $\gamma=10^{-10}$.
We observe that a value of $\tau\omega=1$ or $\tau\omega=0.5$ leads to a maximal error in the one-body density of the order of $10^{-1}$ with respect to the maximal density.
When employing $\tau\omega=0.05$ instead, this error decreases by two orders of magnitude.
Smaller values such as $\tau\omega=0.005$ lead to an error in the density of the same order of magnitude.
Therefore, we use the value of $\tau\omega=5\cdot 10^{-2}$ in Sec.~\ref{sec:ho} since smaller values $\tau\omega$ do not improve the accuracy of the method while leading to a higher computational effort due to more frequent evaluations of the pruning criterion.

%
\end{document}